\documentclass[12pt]{article}
\usepackage{epsfig, amsmath, amssymb}
\usepackage{graphicx,psfrag}
\usepackage{xcolor} 
\newcommand{\nn}{\nonumber}
\newcommand{\be}{\begin{equation}}
\newcommand{\ee}{\end{equation}}
\newcommand{\ben}{\begin{equation}}
\newcommand{\een}{\end{equation}}
\newcommand{\bea}{\begin{eqnarray}}
\newcommand{\eea}{\end{eqnarray}}
\newcommand{\bA}{\begin{array}}
\newcommand{\eA}{\end{array}}
\newcommand{\bc}{\begin{center}}
\newcommand{\ec}{\end{center}}
\newcommand{\al}{\alpha}

\newcommand{\ra}{\rightarrow}
\newcommand{\del}{\partial}

\newcommand{\ie}{{\it i.e.}}
\newcommand{\eg}{{\it e.g.}}

\newcommand{\lan}{\langle}
\newcommand{\ran}{\rangle}

\begin{document}


\begin{titlepage}

\bc

\hfill 

\vspace{25mm}


{\Huge Cosmologies, singularities and \\ [2mm]
  quantum extremal surfaces} 
\vspace{16mm}

{\large  Kaberi Goswami,\ \ K.~Narayan,\ \ Hitesh K. Saini} \\
\vspace{3mm}
{\small \it Chennai Mathematical Institute, \\}
{\small \it SIPCOT IT Park, Siruseri 603103, India.\\}

\ec
\vspace{30mm}

\begin{abstract}
  Following arXiv:2012.07351 [hep-th], we study quantum extremal
  surfaces in various families of cosmologies with Big-Crunch
  singularities, by extremizing the generalized entropy in
  2-dimensional backgrounds which can be thought of as arising from
  dimensional reduction.  Focussing first on the isotropic $AdS$
  Kasner case, introducing a spatial regulator enables relating the
  locations in time of the quantum extremal surface and the
  observer. This shows that the quantum extremal surface lags behind
  the observer location. A potential island-like region, upon
  analysing more closely near the island boundary, turns out to be
  inconsistent. Similar results arise for other holographic
  cosmologies. We then study certain families of null Kasner
  singularities where we find that the quantum extremal surface can
  reach the near singularity region although the on-shell generalized
  entropy is generically singular.  We also study other cosmologies
  including de Sitter (Poincare slicing) and FRW cosmologies under
  certain conditions.
\end{abstract}

\end{titlepage}

{\tiny 
\begin{tableofcontents}
\end{tableofcontents}
}


\section{Introduction}

In \cite{Manu:2020tty}, we studied aspects of entanglement and quantum
extremal surfaces (QES) in various families of holographic spacetimes
exhibiting cosmological singularities.  This is inspired by the
exciting discoveries made recently on the black hole information
paradox
\cite{Penington:2019npb,Almheiri:2019psf,Almheiri:2019hni,Penington:2019kki,Almheiri:2019qdq},
unravelled via the study of entanglement, quantum extremal surfaces
and islands: by now there is a large body of literature on various
aspects of these issues, reviewed in \eg\
\cite{Almheiri:2020cfm,Raju:2020smc,Chen:2021lnq,Kibe:2021gtw}. Quantum
extremal surfaces are extrema of the generalized entropy
\cite{Faulkner:2013ana,Engelhardt:2014gca} obtained by incorporating
the bulk entanglement entropy of matter alongwith the classical area
of the entangling RT/HRT surface
\cite{Ryu:2006bv}-\cite{Rangamani:2016dms}. These lead to various new
insights on black holes. Explicit calculation is possible in effective
2-dimensional models where the bulk entanglement entropy can be
studied through 2-dim CFT techniques.

It is interesting to ask if quantum extremal surfaces might be used to
probe cosmological, Big-Crunch or -Bang, singularities. While the
vicinity of the singularity is expected to be rife with severe
stringy/quantum gravity effects, one might hope to gain some insight
into how these extremal surfaces probe such singularities.  Some
interesting recent work on QES and cosmologies
appears in \cite{Chen:2020tes,Hartman:2020khs} and also
\eg\ \cite{Krishnan:2020fer}-\cite{Aguilar-Gutierrez:2021bns}.

The investigations in \cite{Manu:2020tty} pertained to various
Big-Crunch singularities, in particular the isotropic $AdS$ Kasner
spacetime. These spacetimes have no horizons and no significant
entropy, so they are somewhat unlike black hole horizons. Further, we
are considering closed universes with no entanglement with
``elsewhere'' (\eg\ other universes).  Part of the goal here is to
gain some understanding of how quantum extremal surfaces probe such
spacetime singularities in closed universes with no horizons and no
entanglement with regions external to these universes. The
time-dependence implies that the classical extremal RT/HRT surface
dips into the bulk radial and as well as time directions. Explicitly
analysing the extremization equations in the semiclassical region far
from the singularity can be carried out in detail: we find the surface
bends in the direction away from the singularity. In the 2-dim
cosmologies \cite{Bhattacharya:2020qil} obtained by dimensional
reduction of these and other singularities, quantum extremal surfaces
can be studied by extremizing the generalized entropy, with the bulk
matter taken to be in the ground state (which is reasonable in the
semiclassical region far from the singularity). The resulting
extremization shows the quantum extremal surfaces to always be driven
to the semiclassical region far from the singularity. In
sec.~\ref{sec:rev}, we review the analysis in \cite{Manu:2020tty}. The
2-dim dilaton gravity theories in these cases are somewhat more
complicated than Jackiw-Teitelboim gravity and are not ``near JT'' in
essential ways. The cosmological solutions here are sourced by an
extra scalar which descends from the scalar in the higher dimensional
theory. These theories capture a subset of the observables of the
higher dimensional theory and so are best regarded as models of
``effective holography'' \cite{Narayan:2020pyj}, UV-incomplete in
totality but adequate for capturing various aspects including
entanglement. Since the quantum extremal surfaces are driven to the
semiclassical region far from the singularity, the approximation of
using the 2-dimensional theory is consistent and the other higher
dimensional modes do not make any significant contribution.

In this paper, we continue our investigations there and develop them
further: wherever possible we look for quantum extremal surfaces
spacelike-separated from the observer location. We first do a careful
study of QES focussing on $AdS$ Kasner singularities
(sec.~\ref{sec:AdSKreg}), by introducing a spatial regulator.  This
enables relating the locations in time of the observer on the
holographic boundary and the QES with the bulk matter central charge
and the regulator.  In the semiclassical region, this shows that the
quantum extremal surface lags behind the observer location (in the
direction away from the singularity).  A potential island-like region,
upon analysing in detail near the island boundary, turns out to be
inconsistent.  We then extend this to more general singularities
admitting a holographic interpretation, which exhibit similar
behaviour.  In sec.~\ref{sec:null}, we study certain families of null
Kasner Big-Crunch singularities: these exhibit a certain
``holomorphy'' due to special properties of null backgrounds. Further
they are also distinct in the behaviour of the QES, which now can
reach the singularity (although the generalized entropy continues to
be singular). We then discuss aspects of 2-dimensional effective
theories involving dimensional reduction of other cosmologies in
sec.~\ref{sec:dSFRW}, including de Sitter space (Poincare slicing) and
FRW cosmologies under certain conditions. Sec.~\ref{sec:Disc} contains
some conclusions.  Some details appear in two Appendices.

\section{Review: Big-Crunches \& quantum extremal surfaces}\label{sec:rev}

There is a long history of studying cosmological singularities in
string theory and holography: see \cite{Bhattacharya:2020qil} for a
partial list of references in this regard, and
\eg\ \cite{Craps:2006yb,Burgess:2011fa} for reviews of cosmological
singularities in string theory.
In \cite{Manu:2020tty}, various families of cosmological spacetimes
with spacelike Big-Crunch singularities were considered: the higher
dimensional space and its reduction ansatz are of the form
\be\label{redux+Weyl}
ds^2_D = g^{(2)}_{\mu\nu} dx^\mu dx^\nu + \phi^{2\over d_i} d\sigma_{d_i}^2\ ;
\qquad\quad g_{\mu\nu}=\phi^{{d_i-1\over d_i}} g^{(2)}_{\mu\nu}\ ,
\qquad D=d_i+2\ .
\ee
$d_i$ is the dimension of the transverse space.
The Weyl transformation from $g^{(2)}_{\mu\nu}$ to the 2-dim metric
$g_{\mu\nu}$ ensures that the dilaton kinetic energy vanishes and
the action is
\be\label{actionXPsiU}
S= {1\over 16\pi G_2} \int d^2x\sqrt{-g}\, \Big(\phi\mathcal{R}
- U(\phi,\Psi) -\frac{1}{2} \phi (\partial\Psi)^2 \Big)\ ,
\ee
The dilaton potential $U(\phi,\Psi)$ potentially couples the dilaton
$\phi$ to $\Psi$.\ Certain aspects of generic dilaton gravity theories
of this kind (and these 2-dim cosmological backgrounds), dimensional
reduction and holography were discussed in \cite{Narayan:2020pyj} (see
also \cite{Grumiller:2021cwg}): these theories are more complicated
than JT gravity and are not ``near JT''. they capture a subset of the
observables of the higher dimensional theory and so are best regarded
as UV-incomplete models of ``effective holography''.  There is
nontrivial dynamics in the theory (\ref{actionXPsiU}) driven by the
extra scalar $\Psi$ which descends from the scalar in the higher
dimensional theory. In particular there are nontrivial cosmological
singularity solutions here, which were analysed in
\cite{Bhattacharya:2020qil}. See Appendix~\ref{App:2dgES} for some
details. The power-law scaling ansatze for the 2-dim fields and the
corresponding higher dimensional spacetimes are
\be\label{phie^fPsi-ansatz}
\phi=t^kr^m,\quad e^f=t^ar^b,\quad e^\Psi=t^\al r^\beta \quad\ra\quad
ds_D^2 = {e^f\over \phi^{(d_i-1)/d_i}}\big(-dt^2+dr^2\big)+\phi^{2/d_i}dx_i^2\ .
\ee
The universality (\ref{univSing}) implies that $k=1$.
Note that $r=0$ is the asymptotic (holographic) boundary. The
equations of motion (\ref{2dimseom-EMD1-Psi}) then lead to algebraic
relations between the various exponents above, which can then be
solved for, leading to nontrivial families of cosmological solutions
\cite{Bhattacharya:2020qil}.
A prototypical example is $AdS$ Kasner and its reduction to 2-dimensions,
\bea\label{AdSDK-2d}
&& U=2\Lambda\phi^{1/d_i}\,,\quad \Lambda=-{1\over 2}\,d_i(d_i+1)\ ,\qquad
p={1\over d_i}\ , \quad
\al = \sqrt{{2(d_i-1)\over d_i}} , \nn\\
&&  ds^2 = {R^2\over r^2} (-dt^2 + dr^2) + {t^{2p}\,R^2\over r^2} dx_i^2\,,
\qquad e^\Psi = t^\al\,,\qquad d_ip^2=1-{1\over 2}\al^2\ , \nn\\ [1mm]
\ra\ \ && \phi={t\,R^{d_i}\over r^{d_i}}\,,\qquad
ds^2={t^{(d_i-1)/d_i}\,R^{d_i+1}\over r^{d_i+1}}(-dt^2+dr^2)\,,\qquad
  e^\Psi=t^{\sqrt{2(d_i-1)/d_i}}\ .\quad
\eea
$R$ is the $AdS$ length scale. We are suppressing an implicit Kasner
scale $t_K$: \eg\ $t^{2p}\ra (t/t_K)^{2p}$. We will reinstate this as
required.

The higher dimensional spacetimes and their dual field theories were
in fact studied long back in \cite{Das:2006dz}-\cite{Awad:2008jf} as
certain kinds of time-dependent deformations of $AdS/CFT$ with the
hope of gaining insights via gauge/gravity duality into cosmological
(Big-Bang or -Crunch) singularities: some aspects of these were
reviewed in \cite{Bhattacharya:2020qil}. See also
\cite{Engelhardt:2014mea}-\cite{Engelhardt:2016kqb} for further
investigations on some of these. While the bulk spacetime develops a
cosmological Big-Crunch (or -Bang) singularity and breaks down, the
holographic dual field theory (in the $AdS_5$ case), living on a space
that itself crunches, is subject to a severe time-dependent gauge
coupling $g_{YM}^2=e^\Psi$ and may be hoped to provide insight into
the dual dynamics. In this case the scalar $\Psi$ controls the
gauge/string coupling. Generically it was found by analysing at weak
coupling that the gauge theory response also ends up appearing
singular \cite{Awad:2008jf}: however null singularities appear
better-behaved admitting weakly coupled CFT descriptions in certain
variables \cite{Das:2006pw}\ (a string worldsheet analysis in related
null Kasner singularities appears in \cite{Madhu:2009jh}). There is a
large family of such backgrounds exhibiting cosmological singularities
found long back: in these the deformations of the metric and string
dilaton $\Psi$ are constrained, suggesting that the dual CFT state is
likely nontrivial, with nontrivial non-generic initial conditions
required to create Big-Crunch singularities which are perhaps
qualitatively different from black holes (note that generic severe
time-dependent deformations on the vacuum state are expected to
thermalize on long timescales, dual to black hole formation in the
bulk). Some of these backgrounds have the technical feature of spatial
isotropy which allows studying these backgrounds from a possibly
simpler perspective, by carrying out a dimensional reduction on the
spatial directions with the ansatz (\ref{redux+Weyl}). This enables
the 2-dim dilaton gravity perspective (\ref{actionXPsiU}) formulated
in \cite{Bhattacharya:2020qil}, and also helps uncover new cosmologies
of the form (\ref{phie^fPsi-ansatz}) including ones with
nonrelativistic (hyperscaling violating Lifshitz) asymptotics,
reviewed briefly in Appendix~\ref{App:2dgES}.

\begin{figure}[h] 
\hspace{1pc}
\includegraphics[width=15pc]{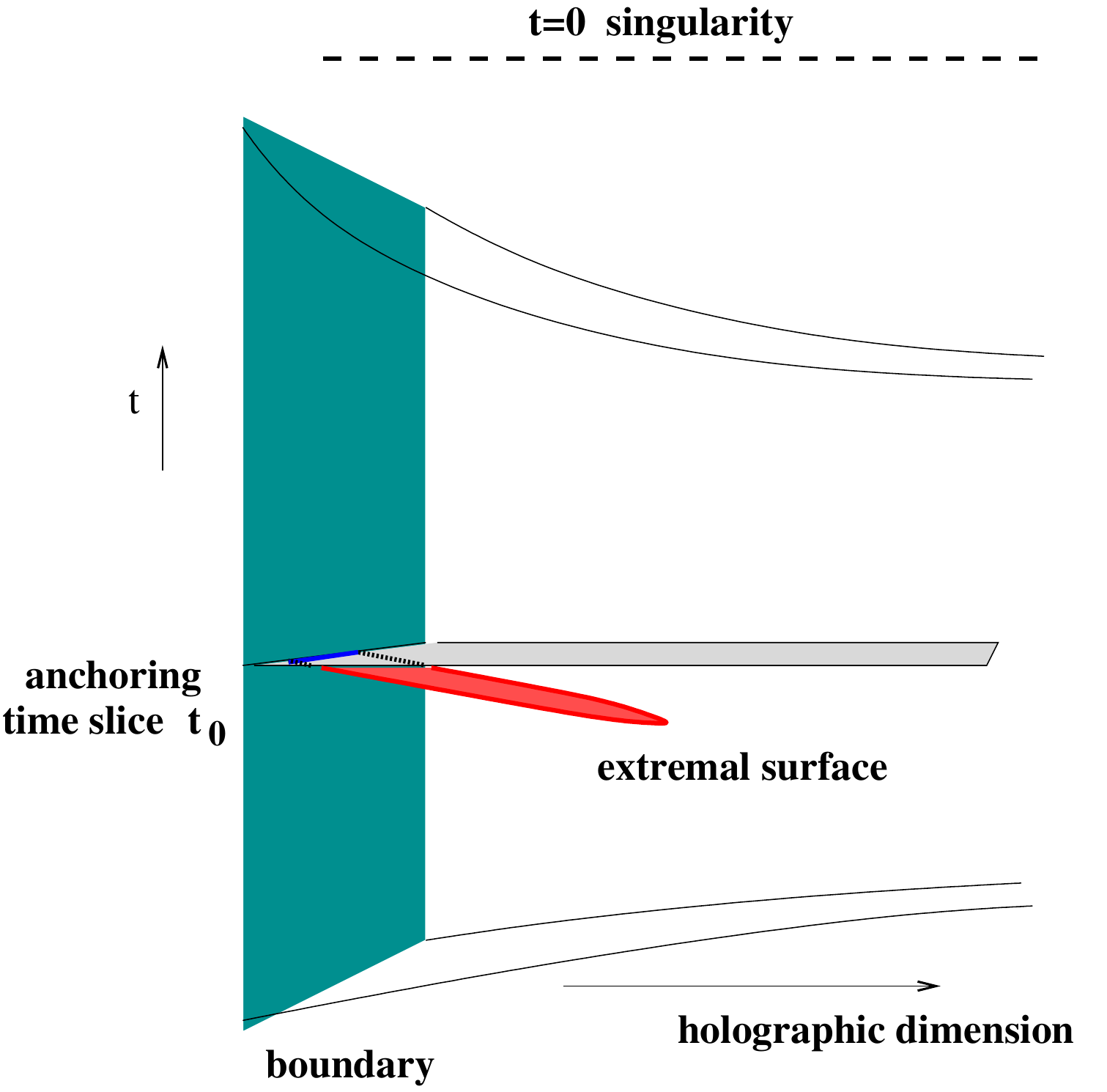}
\hspace{2pc}
\begin{minipage}[b]{19pc}
\caption{{ \label{cosRT1}
\footnotesize{
      Cartoon of extremal surfaces in $AdS$ Kasner spacetime, 
      anchored on a boundary time slice $t_0$ (extended as the grey
      horizontal plane in the bulk). The extremal surface (red) bends
      away from the singularity at $t=0$ (dotted line), \ie\ $t_*>t_0$,\
      with $(t_*,r_*)$ the turning point.    \newline\newline
}}}
\end{minipage}
\end{figure}
Extremal surfaces can be studied as codim-2 surface probes of these
cosmological spacetimes. This is reliable if the surface is anchored on
a boundary subregion in the semiclassical region far from the singularity
where stringy or quantum gravity effects are not large. This can be
analysed in great detail as in \cite{Manu:2020tty}. The time-dependence
of the cosmology implies that the RT/HRT surface dips into the time
direction also, besides the radial (holographic) direction.
The resulting picture (focussing on strip-shaped subregions consistent
with the symmetries here) is as in Figure~\ref{cosRT1}.
The surface is parametrized as $(t(r),x(r))$ stretching in all $x_i$
directions except $x\in \{x_i\}$ which represents the width (size)
direction of the strip with $\Delta x=l$. The anchoring time slice is
$t(0)=t_0$: some details appear in Appendix~\ref{App:2dgES}.
The extremization of the time function $t(r)$ is more complicated due
to the time-dependence and gives a second order nonlinear differential
equation for $t(r)$. In the semiclassical region, we expect the
time-dependence to be mild so that ${dt\over dr}\equiv t'\ll 1$. This
leads to a slightly simpler, but still nonlinear, equation, which
however can be shown to admit power-series solutions,
\be\label{t(r)Expn}
t(r) = t_0 + \sum_n c_n r^n\,, \quad c_n\sim {1\over t_0^\#}
\qquad\Rightarrow\quad t_*>t_0\ ,
\ee
which can then be shown to satisfy $t_*\equiv t(r_*)>t_0$, in other
words, the surface bends in the direction away from the singularity.
This is straightforward to see (although involved) in the regime of
small subregion width (where $A={t_*\over r_*^2}\gtrsim {1\over t_0^2}$).
The analysis is a little more delicate in the IR limit where the
subregion becomes the whole space (and $A\ra 0$): here we find
\be
r_*\ra\infty\,,\quad t_0\ra\infty\,,\quad {t_0\over r_*}\lesssim 1;
\qquad t_*\gtrsim t_0\ .
\ee
Thus the RT/HRT surface is driven to the region far from the singularity
(see also \cite{Engelhardt:2013jda} for similar observations in a
different context): in the IR limit this effectively means infinitely
far from the singularity since $t_*\gtrsim t_0\ra\infty$.

Another notable example with similar behaviour is \cite{Hartman:2013qma},
where the Hartman-Maldacena surfaces exhibit a limiting surface in the
black hole interior.

\medskip

\noindent \underline{{\bf Quantum extremal surfaces}}

Now we briefly review the discussion of quantum extremal surfaces in
\cite{Manu:2020tty}.
Quantum extremal surfaces are extrema of the generalized entropy
$S_{gen} = S_{cl}+S_{bulk}$, the leading classical term being the area
of the extremal surface while the second term is the entropy of the
bulk matter in the region enclosed by the extremal surface and the
boundary. In 2-dim theories, the bulk entropy can be calculated by
using 2-dim CFT techniques. For instance, if the bulk matter is
approximated by a CFT in a curved space and is taken to be in the
ground state, then the bulk entropy can be obtained by a
generalization of the Calabrese-Cardy replica formulation
\cite{Calabrese:2004eu,Calabrese:2009qy} for a single spacetime
interval $\Delta^2$, giving (see Appendix~\ref{App:EE2dCFT})
\be\label{Sgen0}
S_{gen} = {\phi\over 4G} +
{c\over 12}\log \big(\Delta^2 e^f|_{(t,r)} e^f|_{(t_0,r_0)}\big) ;\qquad
1\ll c\ll {1\over G}\ .
\ee
The last condition arises from requiring that the bulk matter entropy
is non-negligible but not so large as to destabilize the leading
classical area contribution.
The 2-dim space is of the form $ds^2=e^f\eta_{\mu\nu}dx^\mu dx^\nu$ and
the Weyl factors above arise from the conformal transformation of
the twist operator 2-point function in the replica formulation, the
twist operators located at the endpoints of the interval in question
(between the boundary and the extremal surface).

As a simple time-independent example consider the 2-dim dilaton-gravity
background obtained from the dimensional reduction (\ref{redux+Weyl}) of
$AdS_{d_i+2}$ with metric in the Poincare slicing\
$ds^2_{AdS_{d_i+2}}={R^2\over r^2} (-dt^2+dr^2)+{R^2\over r^2} dx_i^2$\
($R$ is the $AdS$ scale). Some aspects of such generic 2-dim
dilaton gravity theories have been discussed in \cite{Narayan:2020pyj}.
This 2-dim background, the corresponding generalized entropy (\ref{Sgen0})
and its extremization give
\bea\label{SgenAdSDred}
\phi = {R^{d_i}\over r^{d_i}}\ ,\ \ &&
ds^2 = {R^{d_i+1}\over r^{d_i+1}} (-dt^2+dr^2)\ ,\\
S_{gen} = {\phi_r\over 4G}\,{R^{d_i}\over r^{d_i}}
+ {c\over 12} \log \left( {r^2/\epsilon_{UV}^2\over (r/R)^{d_i+1}} \right)\ \ 
&\Rightarrow &\ \ 
\del_rS_{gen} = -{d_i\phi_rR^{d_i}\over 4G\, r^{d_i+1}}
- {c\over 6}\,\Big({d_i-1\over 2}\Big)\,{1\over r} = 0\ . \nonumber
\eea
(We have written $S_{bulk}$ using the rules of boundary CFT since the
effective space is the half-line with one end of the interval at the
boundary $r=0$: see Appendix~\ref{App:EE2dCFT}. A useful resource for
QES calculations in time-independent cases is \cite{MahajanTalk}.)\ 
We see that both terms are of the same sign since $c>0$ and $d_i>1$.
Thus the solution is\ $r\equiv r_*\ra\infty$ for the location of the QES:
this leads to the entire Poincare wedge which is the expected answer
(also in the higher dimensional point $AdS_D$ when the subsystem
becomes the whole space).
Thus in this case, there are no islands, \ie\ regions disconnected from the
boundary defined \eg\ by a finite location of the quantum extremal
surface\footnote{\label{IslAMM}
  In \eg\ \cite{Almheiri:2019yqk}, a flat non-gravitating (bath)
  region was appended beyond the boundary $r=0$ of an $AdS_2$ region,
  giving the generalized entropy\
$S_{gen}\sim {\phi_r\over 4G}{1\over r} + {c\over 6}\log ((r+r')^2\,{1\over r})$.
The interval in question has endpoints $r\in AdS_2$ and $r'$ in the
flat space region beyond the boundary: the warp factor at the $r'$ end
does not contribute since it is trivial in that flat region.
Both $r, r'>0$ in this parametrization: the space is not a half-line
now.  Setting $r'\sim 0$ for simplicity and extremizing gives\
$-{\phi_r\over 4G}{1\over r^2} + {c\over 6}{1\over r} = 0$\,: the 
competition between the two terms leads to a finite value
$r_*\sim {\phi_r\over Gc}$ for the QES location, \ie\ an island.}.

One way to understand this is in terms of the violation of the
Bekenstein bound, as discussed in \cite{Hartman:2020khs}: if the
classical dilatonic term is overpowered by the subleading bulk
entropy contribution, we may expect islands. To see this, note that
(\ref{SgenAdSDred}) can be recast as
\be\label{noIsl}
S_{gen} = {\phi\over 4G} + {c\over 12}\,{d_i-1\over d_i}\,\log\phi\ ,
\ee
with a relative plus sign in the two contributions, retaining
only terms relevant for extremization. As long as $\phi$ is not too
small, the bulk entropy term scaling as $\log\phi$ is subdominant to
the classical area term scaling as $\phi$. If we entangle the bulk
matter with ``elsewhere'' then $S_{bulk}$ could increase possibly
leading to nontrivial competition with the classical area term and
thereby islands, as is the case in \cite{Almheiri:2019yqk}
and in various cases in \cite{Hartman:2020khs}.

Now we will study quantum extremal surfaces in the 2-dim cosmological
backgrounds reviewed earlier. We focus first on the 2-dim cosmology
obtained by reduction of the $AdS_D$ Kasner spacetime (\ref{AdSDK-2d}),
restricting attention to the observer at the holographic boundary at
$r=0$.
We carry out the extremization in the reliable semiclassical region far
from the singularity at $t=0$: the observer is at $(t_0,0)$.
Assuming for simplicity that the QES lies on the same time slice as
the observer \ie\ $t=t_0$, equivalently that the QES is maximally
spacelike separated from the observer, it turns out that (\ref{Sgen0})
can be recast as
\bea\label{SgenAdSKas-t=t0}
t=t_0:\qquad\quad
S_{gen} = {\phi\over 4G} + {c\over 6}\,{d_i-1\over d_i} \log\phi\ ,&& \quad
\phi={t\over r^{d_i}}\ ,\nn \\ [2mm]
\del_rS_{gen} \sim
-\left( {\phi_r\over 4G}\,{d_i\,t\over r^{d_i+1}} + {c\over 12} {d_i-1\over r}
\right) = 0 ,&&
\del_tS_{gen} \sim
{\phi_r\over 4G}\,{1\over r^{d_i}} + {c\over 12}\,{d_i-1\over d_i\,t} = 0\ .\ \
\eea
Since $c>0$ and $d_i>1$, both contributions in both derivative
expressions appear with the same sign. Note also that in this entire
discussion, we are on one side (the past) of the singularity at $t=0$,
so the range of the time variable is\ $t\equiv |t|\geq 0$.\ Then it is
clear that the only QES solution $(t_*,r_*)$ to extremization is
\cite{Manu:2020tty}
\be\label{AdSKas-t*r*}
t\sim t_0\ ,\qquad r\equiv r_*\ra\infty\ ,\qquad t\equiv t_*\ra\infty\ ;
\qquad t_* \lesssim r_*\ ,
\ee
\ie\ the quantum extremal surface is driven to the semiclassical
region, infinitely far from the singularity at $t=0$. A more general
analysis vindicates this. Further, in this semiclassical region the
dilaton is not too small so there are no islands here since the
Bekenstein bound is not violated, as in (\ref{noIsl}).

\section{$AdS$ Kasner, quantum extremal surfaces, regulated}\label{sec:AdSKreg}

In what follows we will study various 2-dim backgrounds given by the
dilaton $\phi$ and the 2-dim metric $e^f$ and analyse quantum extremal
surfaces obtained from the extremization of the generalized entropy
(\ref{Sgen0}): in general these are of the form
\bea\label{Sgen1}
&& \qquad\qquad
S_{gen}={\phi\over 4G}+{c\over 12}\log(\Delta^2\,e^f|_{(t,r)})\ ,
\qquad\qquad \Delta^2=r^2-(t-t_0)^2\,,\nonumber\\  [1mm]
&&   {\del_r\phi\over 4G} + {c\over 6}\,{r\over\Delta^2}
   + {c\over 12}\del_rf = 0\,,
\qquad
{\del_t\phi\over 4G} - {c\over 6}\,{t-t_0\over\Delta^2} +
{c\over 12} \del_tf = 0\ ,
\eea
where we have retained only terms relevant for extremization.
These are all spaces with a holographic boundary so we are using
the corresponding expression for the generalized entropy in
Appendix~\ref{App:EE2dCFT}.

We would like to understand the dependence of the quantum extremal
surface ($t_*,r_*)$ on the observer location $(t_0,r_0)\equiv (t_0,0)$\,:
we will focus on the observer at the holographic boundary $r=0$.\
Here we study the $AdS$ Kasner case: we will put back the $AdS$ scale
$R$ and the Kasner scale $t_K$ in (\ref{AdSDK-2d}) so the lengthscales
are manifest. Then the dilaton and 2-dim metric become
\be\label{AdSK-RtK}
\phi={t/t_K\,\over (r/R)^{d_i}}\,,\qquad
ds^2={(t/t_K)^{(d_i-1)/d_i}\over (r/R)^{d_i+1}}(-dt^2+dr^2)\ .
\ee
Towards understanding quantum extremal surfaces, let us study
(\ref{Sgen1}) with the scales put in explicitly as in (\ref{AdSK-RtK}).
If we assume $t=t_0$, we obtain (\ref{SgenAdSKas-t=t0}),
(\ref{AdSKas-t*r*}), which are structurally similar to the $AdS$ case
(\ref{SgenAdSDred}). We will instead attempt solving for $t$ as a
function of $t_0$. Then the extremization equations are
(introducing $\phi_r$ as bookkeeping for now)
\be\label{ExtSgen-r0t0}
{c\over 6}\,{r\over \Delta^2} =
{\phi_r\over 4G}\,{d_i\,t/t_K\over \,r^{d_i+1}/R^{d_i}}
+ {c\over 12}\,{d_i+1\over r}\,,\qquad\quad
{c\over 6}\,{t-t_0\over \Delta^2} =
{\phi_r\over 4G}\,{1/t_K\over \,r^{d_i}/R^{d_i}}
+ {c\over 12}\,{d_i-1\over d_i\,t}\ .
\ee
Note that each term now has dimensions of inverse length manifestly.
In the parametrization of these cosmologies (\ref{AdSDK-2d}), the
singularity is at $t=0$: regarding this as a Big-Crunch, we take
the time coordinate $t$ to represent $|t|$ so that $t>0$ in our
entire discussion.

We require that the QES is spacelike-separated from the observer,
consistent with the interpretation of these extremal surfaces as
holographic entanglement.  This implies
\be\label{t>t0}
\qquad \Delta^2 > 0\quad\Rightarrow\quad t_*>t_0\ ,\qquad\qquad\qquad
      [\Delta^2=r^2-(\Delta t)^2]
\ee
from the $t$-equation in (\ref{ExtSgen-r0t0}). This means that the
QES always lags behind the observer, in the direction away from the
singularity ($t=0$).

Let us now look in more detail at QES solutions near the semiclassical
solution (\ref{AdSKas-t*r*}), where $\Delta t\sim 0$ and
$r, t\ra\infty$. Let us first rewrite the $r$-extremization equation
in (\ref{ExtSgen-r0t0}) as
\be\label{rExt-reg}
{3\phi_r\over Gc}{d_i\,t/t_K\over \,r^{d_i+1}/R^{d_i}}
+ \Big({d_i+1\over r} - {2r\over\Delta^2}\Big)
= {3\phi_r\over Gc}{d_i\,t/t_K\over \,r^{d_i+1}/R^{d_i}}
+ {d_i+1\over r} \Big( {{d_i-1\over d_i+1} r^2 - (\Delta t)^2
  \over r^2-(\Delta t)^2} \Big) = 0
\ee
As long as $\Delta t$ is small, \ie\ $\Delta^2\sim r^2$, the second
term is positive: thus both terms are positive, the only solution
to this being $r\equiv r_*\ra\infty$. This is very similar to the
time-independent $AdS$ case in (\ref{SgenAdSDred}), giving the
entire Poincare wedge as the entanglement wedge: there are no islands.

\begin{figure}[h] 
\hspace{2pc}
\includegraphics[width=11pc]{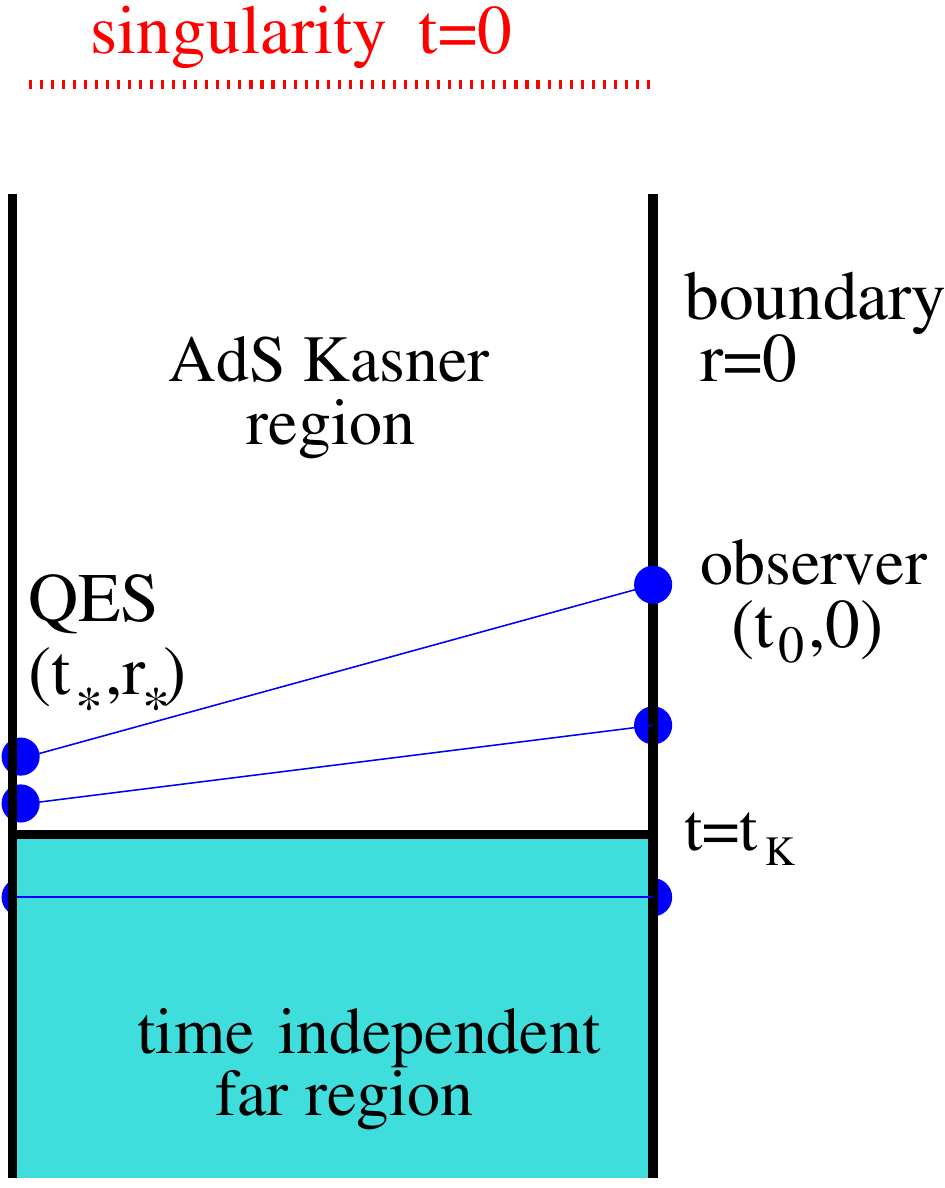}
\hspace{3pc}
\begin{minipage}[b]{21pc}
\caption{{ \label{cosQES3}
    \footnotesize{Cartoon of the 2-dim $AdS$ Kasner geometry
      (singularity at $t=0$), the holographic boundary at $r=0$ and
      the QES at $(t_*,r_*)$, with a time-independent $AdS$ space
      appended for $t>t_K$. The boundary observer $(t_0,0)$ moves in
      time from the time-independent region to the $AdS$ Kasner region.
      The QES lags behind in time, \ie\ $t_*>t_0$, when $t_0$ is in
      the Kasner region.
      \newline \newline 
}}}
\end{minipage}
\end{figure}

Analysing the $t$-extremization equation is rendered tricky with
$r_*\ra\infty$ strictly. Towards obtaining insight into the $t_0$
dependence of $t_*$, let us regulate as $r_*=R_c\sim\infty$ with some
large but finite spatial cutoff $R_c$ that represents the boundary
of the entanglement wedge. Then the $t$-equation in (\ref{ExtSgen-r0t0})
becomes
\be\label{tExt-reg}
\qquad{\Delta t\over R_c^2-(\Delta t)^2} =
{1\over 2K_c}         
+ {d_i-1\over 2 d_i\,t}\ ,
\qquad\qquad
{1\over K_c} = {3\phi_r\over Gc} {1/t_K\over \,R_c^{d_i}/R^{d_i}}\ .
\ee
This expression is manifestly satisfied semiclassically as in 
(\ref{AdSKas-t*r*}). Taking these regulated equations as containing
finite terms we can solve for $t_*$\,: with $\Delta t\ll R_c$, we
obtain the approximate regulated expression
\be\label{tExt-reg2}
\qquad  {\Delta t\over R_c^2} \sim
{1\over 2K_c}
+ {d_i-1\over 2 d_i\,t_0}\ ,
\qquad\quad \Delta t = t_*-t_0\ ,
\ee
where we have approximated $\Delta^2\sim R_c^2$ and set $t\sim t_0$ in
the last expression (with $t_0$ large, as in (\ref{AdSKas-t*r*}))\
(there is some similarity with the semiclassical expansion (\ref{t(r)Expn})).
We see that the QES (\ref{tExt-reg2}) lags behind the observer, in
the direction away from the singularity.
We now see that as $t_0$ decreases, $\Delta t$ increases, \ie\ the
lag of the QES is increasing: see the top part of Figure~\ref{cosQES3}
for a heuristic depiction (the lag is exaggerated!).

The on-shell generalized entropy (\ref{Sgen1}) in the semiclassical
regime where $\Delta^2\sim R_c^2$ becomes
\be\label{SgenOS-semicl}
S_{gen}^{o.s.} \sim {\phi_r\over 4G}\,{t_*/t_K\over (R_c/R)^{d_i}} +
{c\over 12} \log \left( {R_c^2\over \epsilon_{UV}^2}\,
{(t_*/t_K)^{(d_i-1)/d_i}\over (R_c/R)^{d_i+1}} \right)\ ,
\ee
with $t_*$ in (\ref{tExt-reg2}).
Since $t_*\gtrsim t_0$ and $R_c$ is large, $S_{gen}^{o.s.}$ is not
dramatically different structurally from the $AdS$ value
(\ref{SgenAdSDred}), without the $t_*/t_K$ factors. In more detail, we see that the on-shell
$AdS$ expression (\ref{SgenAdSDred}) with $r_*=R_c$ and $\phi|_{r_*}=\phi_*$
becomes\
$S^{o.s.}={\phi_*\over 4G} + {c\over 12}\log\big({R^2\over\epsilon_{UV}^2}
({\phi_*\over\phi_r})^{^{(d_i-1)/d_i}}\big)$  so the log
vanishes when its argument becomes $O(1)$, \ie\ when $\phi_*$ is
sufficiently small. At this point, $S^{o.s.}\sim {\phi_*\over 4G}\sim 0$,
in accord with the physical expectation that the $AdS$ ground state
has zero entropy. In this sense the spatial regulator $R_c$ has
physical meaning as the effective physical boundary of the
entanglement wedge, where $\phi_*$ becomes small enough to be comparable
with $({\epsilon_{UV}\over R})^{^{\#}}$. Note that we can recast $S^{o.s.}$
as (\ref{noIsl}) exactly setting\ 
${1\over\phi_r^{(d_i-1)/d_i}} {R^2\over\epsilon_{UV}^2}\sim 1$ thus fixing
$\phi_r$, which can possibly be regarded as renormalizing
${\phi_r\over G}\equiv {1\over G_r}$\ (and rendering $S_{gen}$ finite).
The above expression (\ref{SgenOS-semicl}) is similar when the
$t_*/t_K$ factors are $O(1)$ so the above arguments apply, and the
overall entropy is not appreciable.

As a further check, note that this QES solution vindicates the maximin
property\footnote{In the semiclassical regime, the
  second derivatives\ \ $\del_t^2S_{gen}|_*\sim
-{c\over 12}\,{d_i-1\over d_i\,t_*^2} - {c\over 6}\,{1\over\Delta_*^2}
- {c\over 3}\,{(t_*-t_0)^2\over \Delta_*^4} < 0$\ \ and \ $\del_r^2S_{gen}|_*\sim {\phi_r\over 4G}\,{d_i\,(d_i+1)\,t_*\,R^d_i\over t_k\, R_c^{d_i+2}} + {c\over 12}\,{d_i+1\over R_c^2} + {c\over 6}\,{1\over \Delta_*^2}\,{(1-{2R_c^2\over \Delta_*^2})} >0$\ confirm time-maximization and spatial minimization, with
the regulator $R_c$ finite.}.

Naively it appears that ${\Delta t\over R_c^2} \sim {1\over t_0}$ shows
a growth as $t_0$ decreases. Rewriting (\ref{tExt-reg}) and solving
as a quadratic, taking $\Delta t > 0$, gives
\be
{\Delta t\over R_c} = {1\over R_c} \left( \sqrt{ {1\over
    ({1\over K_c} + {d_i-1\over d_i\,t} )^2} + R_c^2}\ -\
{1\over {1\over K_c} + {d_i-1\over d_i\,t}} \right) \ ,
\ee
showing a slow growth in $\Delta t$ as $t$ decreases, for fixed
regulator $R_c$. Extrapolating and setting $t_0=0$ shows that $t=0$
is not a solution (this can also be seen in (\ref{ExtSgen-r0t0})).

Our analysis is best regarded as valid in the semiclassical regime,
far from the singularity, approximating bulk matter to be in the
ground state.  However perhaps the qualitative feature of the quantum
extremal surfaces and the associated entanglement wedge excluding the
near singularity region (depicted schematically in the top, $AdS$
Kasner, part of Figure~\ref{cosQES3}) will remain as a reliable result
even with better near singularity bulk entropy models.

We note that the $S_{gen}^{o.s.}$ (\ref{SgenOS-semicl}) decreases with
time evolution towards the singularity (this is reminiscent of
\eg\ \cite{Barbon:2015ria,Caputa:2021pad} revealing low complexity in
such singularities). Recasting this semiclassical value in the form
(\ref{noIsl}), we note that as long as $\phi$ is not too small, the
bulk entropy term is subleading to the area term. Thus the Bekenstein
bound is not violated and there are no spatially disconnected islands
of the kind noted in black holes. In some qualitative sense, it is
tempting to regard the excluded near singularity region as a timelike
separated island-like region: it would be interesting to understand
this better.

\subsection{Searching for islands}\label{sec:AdSK-islands}

Looking now at (\ref{rExt-reg}), we see that for
\be\label{islandRegime}
{d_i-1\over d_i+1} r^2 < (\Delta t)^2 < r^2\ ,
\ee
a spacelike-separated island appears to emerge. Unlike the semiclassical
region with $\Delta t\ll r$ (where both terms are the same sign), the
numerator in the term in brackets in (\ref{rExt-reg}) now changes sign
indicating a large but finite $r\sim ({\phi_r\over Gc})^\#$ solution 
leading to a disconnected region: there is some structural similarity
to the discussion in \cite{Almheiri:2019yqk} (see Footnote~\ref{IslAMM}).
Towards exploring this in detail, first, note that the $\del_r$-equation
in (\ref{ExtSgen-r0t0}) can be rewritten as
\be\label{delr-Delta^2}
\qquad
\Delta^2 = {2 r^2\over d_i+1}\, {1\over 1+{d_i\over d_i+1} {t\over K}}\ ,
\qquad\quad   {1\over K} = {3\phi_r\over Gc} {1/t_K\over \,r^{d_i}/R^{d_i}}\ ,
\ee
so
\be\label{Deltat/r-3}
\Delta^2=r^2-(\Delta t)^2\quad\Rightarrow\quad
{\Delta t\over r} = \sqrt{ {{d_i-1\over d_i+1}\
+\ {d_i\over d_i+1} {t\over K} \over
1\ +\ {d_i\over d_i+1} {t\over K} }}\ . 
\ee
The potential island arises at large finite $r$ in (\ref{rExt-reg}) when
\be\label{islandBndry}
(\Delta t)^2\gtrsim {d_i-1\over d_i+1} r^2
\ee
so that $\Delta t$ is not small but in fact scales as $r$ which is
large.
Expanding (\ref{Deltat/r-3}) in the vicinity of (\ref{islandBndry}) gives
\be\label{Deltat/r-3-Expn}
{\Delta t\over r^2} \sim \sqrt{{d_i-1\over d_i+1}}\,{1\over r}
   \left( 1 + {d_i\over d_i^2-1}\,{t\over K} + \ldots \right) .
\ee
Now the $\del_t$-equation (\ref{ExtSgen-r0t0}) as an exact quadratic
can be solved to obtain (choosing $\Delta t>0$)
\be\label{Dtr-tEx-quadr}
{\Delta t\over r} =  \sqrt{ {{d_i^2\over (d_i-1)^2}\,{t^2\over r^2}\over
    ({d_i\over d_i-1}{t\over K} + 1 )^2} + 1}\ -\
{{d_i\over d_i-1}\,{t\over r}\over {d_i\over d_i-1}{t\over K} + 1}\ ,
\ee
with $K$ defined in (\ref{delr-Delta^2}).\ For a nontrivial island-like
solution, this expression for ${\Delta t\over r}$ must match that in
(\ref{Deltat/r-3}) in the vicinity of the island boundary
(\ref{islandBndry}). With ${t\over K}\sim \epsilon$ being small, we
expand and obtain at leading order
\be\label{Isl-leadingterm0}
\sqrt{1+x^2} - x = \sqrt{{d_i-1\over d_i+1}}\quad\xrightarrow{solving}\quad
x\equiv {d_i\over d_i-1}\,{t\over r} = {1\over\sqrt{d_i^2-1}}
\ee
This gives
\be\label{Isl-leadingterm}
\Delta t \gtrsim \sqrt{{d_i-1\over d_i+1}}\ r \sim d_i\, t\ .
\ee
The last condition $\Delta t \gtrsim d_i\,t$ is clearly impossible
with $\Delta t=t-t_0$ for any $d_i>1$.

In addition, using the leading term matching condition
(\ref{Isl-leadingterm}) and expanding (\ref{Dtr-tEx-quadr}) about
the potential island boundary (\ref{islandBndry}) shows that the
first subleading term in ${t\over K}$ is\
${1\over d_i}\sqrt{{d_i+1\over d_i-1}} {t\over K}$\ which does not
match the first subleading term in (\ref{Deltat/r-3-Expn}).

We have been looking for an island-like solution in the vicinity of
the potential island boundary (\ref{islandBndry}) emerging
continuously from the semiclassical region where $r_*\ra\infty$ as
discussed after (\ref{rExt-reg}).  So we require a simultaneous
solution to the extremization equations (\ref{ExtSgen-r0t0}) recast as
(\ref{Deltat/r-3}) and (\ref{Dtr-tEx-quadr}), just inside the island
region. Then at the very least the leading and first subleading terms
in the expansions of (\ref{Deltat/r-3}) and (\ref{Dtr-tEx-quadr}) near
(\ref{islandBndry}) must agree, which is not the case. Thus this
potential island solution is inconsistent.

One could ask if there are nontrivial islands further away, towards
the singularity (although they may not be physically reliable). In
this regard, we can write $\Delta t=t-t_0$ and expand out the $r$- and
$t$-extremization equations (\ref{rExt-reg}), (\ref{tExt-reg})\,: this
leads to two cubic equations in $t$. However, taking them as
simultaneously true (and \eg\ eliminating the $t^3$ term), it appears
that there are no consistent finite $r, t$ solutions to these, \ie\ no
islands.

\subsection{Appending a time-independent far region}

Let us now consider appending the $AdS$ Kasner space with a
time-independent $AdS$ region far from the singularity, joined
at the Kasner scale $t=t_K$. See Figure~\ref{cosQES3}. So we have
$AdS$ Kasner for $t<t_K$ and the time-independent $AdS$ space for
$t>t_K$, \ie\
\bea\label{AdSK-RtK-reg}
&& \phi={t/t_K\,\over (r/R)^{d_i}}\,,\qquad
ds^2={(t/t_K)^{(d_i-1)/d_i}\over (r/R)^{d_i+1}}(-dt^2+dr^2)
\qquad\quad [t<t_K]\ , \nonumber\\
&& \phi={1\over (r/R)^{d_i}}\,,\qquad\ \ \ 
ds^2={1\over (r/R)^{d_i+1}}(-dt^2+dr^2) \qquad\qquad [t>t_K]\ .
\eea
The spaces are joined continuously at $t=t_K$ but the joining is not
smooth.
Now the extremization equations must be analysed separately as the
observer at $t_0$ moves through each region. The generalized entropy
and its extremization (\ref{Sgen1}) applied to the background profiles
(\ref{AdSK-RtK-reg}) in both regions give
\bea
t_0> t_K: &&
{c\over 6}\,{r\over \Delta^2} =
{\phi_r\over 4G}\,{d_i\over \,r^{d_i+1}/R^{d_i}}
+ {c\over 12}\,{d_i+1\over r}\,,\qquad
{c\over 6}\,{t-t_0\over \Delta^2} = 0\ ; \\
t_0< t_K: && {c\over 6}\,{r\over \Delta^2} =
{\phi_r\over 4G}\,{d_i\,t/t_K\over \,r^{d_i+1}/R^{d_i}}
+ {c\over 12}\,{d_i+1\over r}\,,\qquad
{c\over 6}\,{t-t_0\over \Delta^2} =
{\phi_r\over 4G}\,{1/t_K\over \,r^{d_i}/R^{d_i}}
+ {c\over 12}\,{d_i-1\over d_i\,t}\ . \nonumber
\eea
In the time-independent region $t>t_K$ we see it is physically reasonable
to set $t_*=t_0$, \ie\ the QES lies on the same time slice as the observer.
This follows from time-translation invariance in that region at least
for $t_0\gg t_K$ (far from the junction at $t_K$). Since the joining
slice $t_K$ is in the semiclassical region far from the singularity,
it is adequate to use (\ref{tExt-reg2}) with the regulator to study
the time evolution of the QES in the Kasner region. The lagging (or
repulsive) feature of the QES thus begins once the observer transits
into the Kasner region (the sharp joining at $t_k$ implies that the
lag does not evolve smoothly).

To see this in more detail, consider the time $t_0=t_K-\delta t_0$ when
the observer is just entering the Kasner region: then we expect that
the quantum extremal surface is just a little away from the observer
time slice $t_0$. To quantify this, let us compare $\delta t_*$ in 
(\ref{tExt-reg2}) with $\delta t_0$\ (and $K_c$ defined in
(\ref{tExt-reg})): we have
\be
\delta t_0=t_K-t_0>0\ ;\qquad
{\delta t_*\over R_c^2} = {t_*-t_0\over R_c^2} \sim 
{1\over 2K_c} + {d_i-1\over 2d_i\,t_K} \Big( 1 + {\delta t_0\over t_K} \Big) ,
\ee
so that for small $\delta t_0$ \ie\ $t_0\sim t_K$, the quantum extremal
surface ends up being pushed to the time-independent region ($t_*>t_K$).
Of course as the observer moves in time further, the QES enters the
Kasner region as well. To see this further, let us compare the QES
location with the Kasner scale: with $t_0\lesssim t_K$, we
have    
\be
t_*\lesssim t_K \qquad\Rightarrow\qquad
{t_K-t_0\over R_c^2} \gtrsim {1\over 2K_c} + {d_i-1\over 2d_i\,t_0}
\ee
In other words, the quantum extremal surface is within the
Kasner region if the observer is sufficiently further within.
The cross-over of the QES to the Kasner region occurs when $t_*\sim t_K$,
\ie\ when the above inequality is saturated (giving
${t_0-t_K\over R_c^2} \sim - {1\over 2K_c} - {d_i-1\over 2d_i t_K}$).

The model (\ref{AdSK-RtK-reg}) is just meant as a simple toy model for
gaining some insight into the evolution of the quantum extremal
surface as the observer transits from the time-independent far region
into the time-dependent $AdS$ Kasner region towards the
singularity. The existence of the time-independent far region suggests
that one can prepare the initial state as the ground state via a
Euclidean continuation. Putting this on firmer footing is however more
tricky. There is a discontinuity at the $t=t_K$ slice perhaps
reflecting the fact that the Kasner time-dependence does not switch
off at $t_K$: this might imply additional concerns in smooth time
evolution into the Kasner region (without any external energy-momentum
inflow). More detailed analysis of this requires detailed
understanding of the junction conditions for joining up the spacetimes
at $t_K$. Perhaps rather than a sharp time slice at $t_K$, it would be
more physical to find a thickened spacetime region interpolating
smoothly between the time-independent far region and the Kasner
region: then the QES lag is likely to evolve smoothly. We will leave
these questions for the future.

\subsection{More general 2-dim cosmologies, QES, regulated}

In the previous subsections, we studied $AdS$-Kasner cosmologies and
their 2-dim reflections obtained by dimensional reduction
(\ref{AdSDK-2d}), and quantum extremal surfaces.  Now we will extend
this to more general 2-dim cosmologies (\ref{phie^fPsi-ansatz}).
We have the 2-dim dilaton and metric fields of the form
\be\label{genCos}
\phi=t r^m\,,\quad e^f=t^ar^b\,,\qquad a>0,\ \ \ m<0,\ \ \ b<0\ .
\ee
Note that we have taken the time exponent of the dilaton in accord
with the universality (\ref{univSing}) of the near singularity region
found in \cite{Bhattacharya:2020qil}.  We take $a>0$ to simulate a
Big-Crunch singularity at $t=0$. Further we assume $m, b<0$ in accord
with the intuition that the dilaton and the 2-dim metric grow towards
the holographic boundary at $r=0$.

The generalized entropy (\ref{Sgen1}) and its extremization with $r, t$, give
\be\label{ExtSgen-r0t0-gen}
{c\over 6} {r\over\Delta^2}
= {\phi_r\over 4G} {|m| t\over r^{|m|+1}} + {c\over 12} {|b|\over r}\ ,
\qquad
{c\over 6} {t-t_0\over\Delta^2}
= {\phi_r\over 4G} {1\over r^{|m|}} + {c\over 12} {a\over t}\ ,\quad
\ee
analogous to (\ref{ExtSgen-r0t0}), except that we have suppressed
length scales analogous to $R,\, t_K$ here.  Firstly, requiring the
spacelike condition $\Delta^2>0$ implies $t_*>t_0$, analogous to
(\ref{t>t0}): this means the QES lags behind the observer, in the
direction away from the singularity at $t=0$.

As noted already in \cite{Manu:2020tty}, it is clear that the QES
solution to these extremization equations is again of the form
(\ref{AdSKas-t*r*}), \ie\ $r_*\ra\infty,\ t_*\sim t_0\ra\infty$ with
$t_*\lesssim r_*$.  In the vicinity of the semiclassical region,
analogous to the $AdS$ Kasner case (\ref{rExt-reg}) we can recast
the $r$-equation as
\be\label{rExt-reg-gen}
{3\phi_r\over Gc} {|m| t\over r^{|m|+1}}
+ \Big({|b|\over r} - {2r\over\Delta^2}\Big) =
{3\phi_r\over Gc} {|m| t\over r^{|m|+1}}
+ {|b|\over r} \Big({{|b|-2\over |b|}\, r^2 - (\Delta t)^2
  \over r^2-(\Delta t)^2} \Big) = 0\ .
\ee
As in that case, with $\Delta t$ small, \ie\ $\Delta^2\sim r^2$,
both terms are positive and the only solution to this is $r_*\ra\infty$,
giving the entire Poincare wedge as the entanglement wedge: there are
no islands.\
Now, the $t$-equation becomes
\be\label{tExt-reg-gen}
{\Delta t\over R_c^2-(\Delta t)^2} =
{3\phi_r\over 2Gc} {1/t_K\over \,R_c^{d_i}/R^{d_i}} + {d_i-1\over 2 d_i\,t}\ ,
\ee
analogous to (\ref{tExt-reg}).
As before, we are regulating the QES solution as $r_*=R_c\sim\infty$
with some large but finite spatial cutoff $R_c$ representing the
boundary of the entanglement wedge. 
Taking these regulated equations as containing finite terms we can
solve for $t_*$\,, obtaining an approximate regulated expression
analogous to (\ref{tExt-reg2}) after setting $\Delta^2\sim R_c^2$
and $t\sim t_0$. The resulting semiclassical picture is similar to
the discussion in the $AdS$ Kasner case, with the QES lag increasing
as $t_0$ decreases.

Now let us look for island-like solutions in these more general
holographic cosmologies, analogous to Sec.~\ref{sec:AdSK-islands}.
The corresponding island boundary here, analogous to (\ref{islandBndry}), is
\be\label{islandBndry-gen}
(\Delta t)^2\gtrsim {|b|-2\over |b|} r^2\ .
\ee
Analogous to (\ref{Deltat/r-3}) and (\ref{Dtr-tEx-quadr}) in the
$AdS$ Kasner case, we obtain, respectively,
\be\label{Deltat/r-3-gen}
\Delta^2=r^2-(\Delta t)^2\quad\Rightarrow\quad
{\Delta t\over r} = \sqrt{ {{|b|-2\over |b|}\
+\ {|m|\over |b|} {t\over K} \over
1\ +\ {|m|\over |b|} {t\over K} }}\ ,
\qquad {1\over K} = {3\phi_r\over Gc} {1\over r^{|m|}}\ ,
\ee
rearranging (\ref{rExt-reg-gen}), and
\be\label{Dtr-tEx-quadr-gen}
{\Delta t\over r} =  \sqrt{ {{1\over a^2}\,{t^2\over r^2}\over
    ({1\over a}{t\over K} + 1 )^2} + 1}\ -\
{{1\over a}\,{t\over r}\over {1\over a}{t\over K} + 1}\ ,
\ee
from the $\del_t$-equation in (\ref{ExtSgen-r0t0-gen}) regarded as
a quadratic, choosing $\Delta t>0$.

For a nontrivial island-like solution emerging in the vicinity of 
(\ref{islandBndry-gen}), these two expressions for ${\Delta t\over r}$
must match: expanding, the leading order terms give
\be\label{Isl-leadingterm-gen}
x\equiv {1\over a}\,{t\over r}\,:\quad
\sqrt{1+x^2} - x = \sqrt{{|b|-2\over |b|}}\quad\xrightarrow{solving}\quad
{t\over r} = {a\over\sqrt{|b| (|b|-2)}}\ ,
\ee
while matching the first subleading terms requires 
\be
{1\over a\sqrt{|b| (|b|-2)}} \Big( 1 - {1\over \sqrt{|b| (|b|-2) + 1}} \Big)
\,{t\over K} = {|m|/|b|\over \sqrt{|b| (|b|-2)}}\,{t\over K}
\ee
\ie\
\be\label{Isl-subleadingterm-gen}
{a |m|\over |b|} = 1 - {1\over \sqrt{|b| (|b|-2) + 1}}
\ee
For the $AdS$ Kasner values $a={d_i-1\over d_i}\,,\
m=-d_i\,,\ b=-(d_i+1)$, these agree with the conditions obtained in
Sec.~\ref{sec:AdSK-islands}, which were not consistent as we saw.
The condition (\ref{Isl-leadingterm-gen}) gives
$\Delta t = {|b|-2\over a} t$\,: this is impossible in the $AdS$
Kasner case (\ref{Isl-leadingterm}) as we saw.
For the hyperscaling violating cosmologies (\ref{hvL-2d}), this
condition can again be shown to be impossible to satisfy ($a$ takes its
maximum value for $\gamma=0$). The hyperscaling violating Lifshitz
cosmologies in \cite{Bhattacharya:2020qil} require $a=|b|-2\,,\ m=-1$\
(reviewed very briefly after (\ref{hvL-2d})). This gives\
$\Delta t = {|b|-2\over a} t = t$\,,\ which is satisfied for $t_0=0$,
but this is the location of the singularity which is unreliable\
(the condition (\ref{Isl-subleadingterm-gen}) becomes\
${2\over b}={1\over \sqrt{|b| (|b|-2) + 1}}$\ giving $b=-2, a=0$).
Thus overall, these more general holographic cosmologies appear
qualitatively similar to the $AdS$ Kasner case.

The conditions (\ref{genCos}) on the exponents are motivated by the
more general investigations on 2-dimensional cosmologies in
\cite{Bhattacharya:2020qil}.  These investigations employ fairly
general and minimal assumptions on the effective action governing such
cosmological spacetimes: the resulting space of cosmologies is quite
rich, including ones with nonrelativistic (\eg\ hyperscaling violating
Lifshitz) asymptotics and boundary conditions, and they all satisfy
the conditions (\ref{genCos}).  However it would be interesting to
explore the space of such cosmologies, possibly enlarging them
(including those that do not admit reduction to 2-dimensions), towards
understanding the behaviour of quantum extremal surfaces with regard
to the Big-Crunch (-Bang) singularities they may exhibit.

\section{Null cosmologies and quantum extremal surfaces}\label{sec:null}

We consider cosmological spacetimes with null time-dependence in this
section: there are parallels with the discussions in
\cite{Das:2006dz,Das:2006pw,Madhu:2009jh}, as well as
\eg\ \cite{Horowitz:1989bv,Craps:2005wd,Chu:2006pa,Lin:2006ie,Craps:2008bv}.
If we further require that the higher dimensional spacetime admits
dimensional reduction (\ref{redux+Weyl}) to 2-dimensions, this reduces
to a restricted family of 2-dimensional backgrounds of the form
\be\label{nullBgnd}
ds^2=-dx^+dx^-\,,\qquad \phi=\phi(x^+)\,,\qquad \Psi=\Psi(x^+)\ ,
\qquad x^\pm = t\pm r\ .
\ee
The 2-dim metric can always be coordinate-transformed to be flat if
we only have $x^+$-dependence in $\phi, e^f$ in the reduction ansatz
(\ref{redux+Weyl}), leading to the above. The upstairs spacetime
(\ref{phie^fPsi-ansatz}) then is
\be\label{nullUpstairs}
ds^2 = -\phi^{-(d_i-1)/d_i} dx^+dx^- + \phi^{2/d_i} dy_i^2\ ,
\qquad x_i=\{r,y_i\}\ .
\ee
This comprises various higher dimensional backgrounds with null
singularities \eg\
\be\label{nullUpstairs2}
ds^2=(x^+)^a (-dx^+dx^-) + (x^+)^bdy_i^2
\ee
which however are somewhat special, given the restriction to
the 2-dimensional reduction ansatz (\ref{redux+Weyl}): thus it also
does not include the null holographic $AdS$ cosmologies in
\cite{Das:2006dz,Chu:2006pa,Lin:2006ie,Das:2006pw} which are of the form\
$ds^2={R^2\over r^2}[e^{f(x^+)}(-dx^+dx^-+dx_i^2)+dr^2]$\,. There are
qualitative parallels however. The exponents $a,b$ in
(\ref{nullUpstairs2}) are related by the Einstein equations. These
are a bit similar to the null Kasner backgrounds considered in 
\cite{Madhu:2009jh}, except that the 2-dim restriction implies that
$e^f\equiv (x^+)^a$ can be absorbed by redefining the null time
variable $x^+\ra X^+=\int e^f dx^+$. In writing the 2-dim backgrounds
(\ref{nullBgnd}) we have effectively redefined the lightcone variables
$x^\pm$ in this manner. These backgrounds are likely supersymmetric.

Now the equations of motion (\ref{2dimseom-EMD0-Psi}) simplify
tremendously since there is only null-time dependence in the
background ansatze (\ref{nullBgnd}): for instance all nontrivial
contractions of the form $g^{\mu\nu}\del_\mu\Psi\del_\nu\Psi\sim
g^{+-}\del_+\Psi\del_-\Psi$ vanish since there is no $x^-$-dependence.
We also have ${\cal R}=0$ since the 2-dim space is flat. Thus the
equations of motion give
\bea\label{EOMnull}
(++): && -\del_+^2\phi - {\phi\over 2} (\del_+\Psi)^2 = 0\ ;\qquad \\
(\phi):\ \ \ {\del U\over \del\phi} = {\cal R}
- {1\over 2}(\del\Psi)^2 = 0\ ; &&\quad
(\Psi):\ \ \ {\del U\over\del\Psi} = \del_\mu(\phi g^{\mu\nu}\del_\nu\Psi) = 0\ .
\nonumber
\eea
These imply that the dilaton potential is trivial and give a single
nontrivial condition from the $(++)$ equation relating $\phi, \Psi$.
We want to consider a Big-Crunch singularity arising at $x^+=0$ as a
future null singularity, so we take $x^+<0$ in our entire discussion
below. Then
\bea\label{null2d}
&& \phi=(-x^+)^k\,,\quad \Psi=\Psi(x^+)\quad \Rightarrow\quad
(\del_+\Psi)^2 = -2 {\del_+^2\phi\over\phi} = - {2k(k-1)\over (x^+)^2}\ ,
\nonumber\\
&& \Rightarrow\quad 0<k\leq 1\,,\qquad \phi=(-x^+)^k\,,\qquad
e^\Psi = (-x^+)^{\pm\sqrt{2k(1-k)}}\ .
\eea
While $k>0$ gives vanishing dilaton as $x^+\ra 0$, the exponent of
$e^\Psi$ could have either sign. The single $\phi,\Psi$-relation allows
extrapolating $\phi, \Psi$ above to asymptotically constant functions
\ie\ flat space.
This 2-dim background implies the upstairs background
(\ref{nullUpstairs}) with $\phi$ as above: this is of the form
(\ref{nullUpstairs2}) with $a=-{k(d_i-1)\over d_i}$ and $b={2k\over d_i}$\,.
These have\
$R^i{_{+i+}}={k(1-k)\over d_i\,(x^+)^2}$ so tidal forces diverge\
(all curvature invariants vanish due to the null nature of the
backgrounds). To see this in more detail, consider a null geodesic
congruence propagating along $x^+$ with cross-section along some
$y^i$-direction: the geodesic equation then gives
\be
 {dx^+\over d\lambda^2} + \Gamma^+_{++}\Big({dx^+\over d\lambda}\Big)^2 = 0
   \quad\ra\quad  \lambda={(x^+)^{a+1}\over a+1}\ ,
\ee
where $\Gamma^+_{++}={a\over x^+}$\ is the only nonvanishing
$\Gamma^+_{ij}$ component. Solving this leads to the affine parameter
above and the tangent vector becomes
$\xi=\del_\lambda=({dx^+\over d\lambda})\del_+$ so $\xi^+=(x^+)^a$. The
relative acceleration of neighbouring geodesics then is\
$a^M=R^M{_{CDB}}\xi^C\xi^Dn^B$ with $n=n^B\del_B$ the unit normalized
cross-sectional separation vector. Then it can be seen that
$a^i=R^i{_{+i+}}(\xi^+)^2n^i$ so $|a^i|^2$ diverges for all $0<k<1$
leading to diverging tidal forces, somewhat similar to the
corresponding discussion in \cite{Madhu:2009jh}. For $k=1$
the spacetimes (\ref{nullUpstairs}) have all Riemann components
vanishing: these can be recast as $ds^2=-dX^+dx^-+(X^+)^2dy_i^2$ which
can be shown to be flat space in null Milne coordinates (redefining
$Y_i=X^+y_i\,,\ y^-=x^-+y_i^2X^+$).

\begin{figure}[h] 
\hspace{2pc}
\includegraphics[width=11pc]{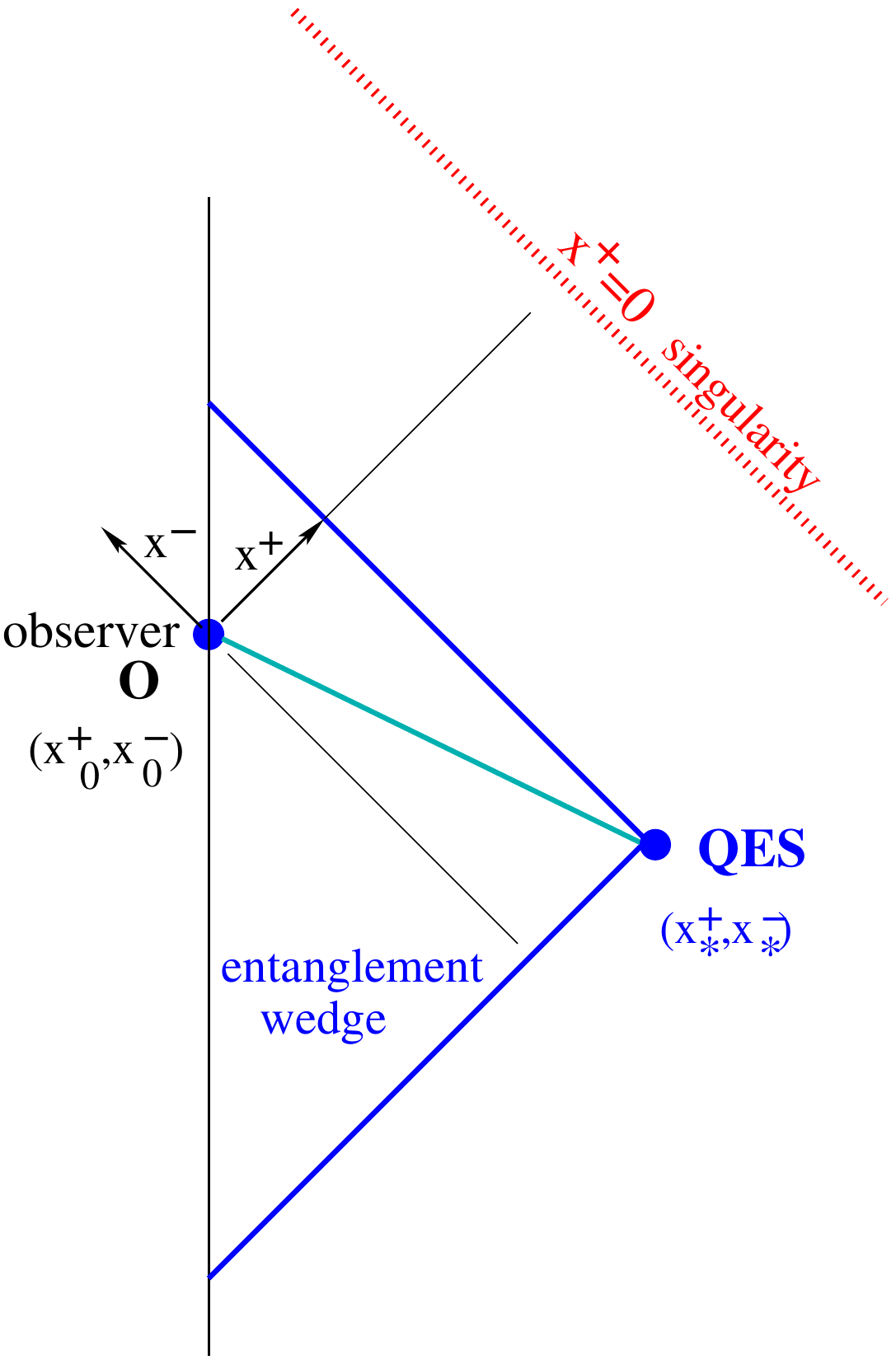}
\hspace{3pc}
\begin{minipage}[b]{20pc}
\caption{{ \label{nullQES}
    \footnotesize{Cartoon of the 2-dim geometry with the 
      null singularity at $x^+=0$, the worldline $(x^+_0,x^-_0)$
      of a  timelike observer (vertical trajectory, representing
      for simplicity a fixed spatial location), and the quantum
      extremal surface at $(x^+_*,x^-_*)$.
      As can be seen, the QES is spacelike separated from the
      observer ($\Delta^2>0$) if $\Delta x^+>0$ and $\Delta x^-\sim-\infty$,\
      and lies towards the singularity in terms of $x^+$-slices.
      The entanglement wedge defined by the QES is shown as the blue
      wedge.
      \newline \newline
      }}}
\end{minipage}
\end{figure}

Now we analyze quantum extremal surfaces. These cosmologies have no
holographic boundary: introducing a bookkeeping $\phi_r$, the
generalized entropy (Appendix~\ref{App:EE2dCFT}) is
\be\label{SgenNull}
S_{gen} = {\phi_r\over 4G} (-x^+)^k + {c\over 6}\log(-\Delta x^+ \Delta x^-)\ ,
\ee
where $\Delta x^\pm=x^\pm-x_0^\pm$ characterizes the spacetime interval
between the observer O and the QES (see Figure~\ref{nullQES}).
Strictly speaking, there is a null Kasner scale $t_N$ here appearing
as $\phi=({-x^+\over t_N})^k$ so $\phi$ is dimensionless: however
since the 2-dim metric is flat in these variables, $t_N$ can be
absorbed into the definition of $\phi_r$ above: so we will suppress
this (unlike the spacelike cases in sec.~\ref{sec:AdSKreg} earlier).\
The extremization with respect to $x^-$ and $x^+$ gives
\be\label{SgenExtNull}
\del_-S_{gen} = {c\over 6} {-\Delta x^+\over -\Delta x^+ \Delta x^-} = 0\,,
\qquad
\del_+S_{gen} = -{\phi_r\over 4G}\, {k\over (-x^+)^{1-k}} + 
    {c\over 6} {\del_+\Delta^2\over \Delta^2} = 0\, .
\ee
With $0<k<1$, the classical extremization ($c=0$) gives $x^+\ra\infty$\,: 
in full, we have
\be\label{nullQESext0}
\Delta^2=-\Delta x^+ \Delta x^->0\,,\quad
\Delta x^-=x^--x_0^-\ra-\infty\,,\quad
       -{1\over (-x^+)^{1-k}} + {2Gc\over 3\phi_rk}\,{1\over x^+-x^+_0} = 0\ ,
\ee
so
\be\label{nullQESext}
\Delta x^+ > 0, \quad
x_*^+=x^+_0 + {2Gc\over 3k\phi_r} (-x_*^+)^{1-k} > x^+_0\ ;
\qquad \Delta x^-<0,\quad x_*^-\ra X_c^-\sim-\infty\ .
\ee
This is best visualized as in Figure~\ref{nullQES}\,: we describe this
further below.
From (\ref{Sgen0}), we have $Gc\ll 1$ so that $x^+\sim x^+_0$ upto
small corrections (with $k\neq 0$). Thus employing perturbation
theory in $Gc$, we obtain
\be\label{nullQESx+}
x_*^+ \sim x^+_0 + {2Gc\over 3k\phi_r} (-x^+_0)^{1-k}\ ,
\ee
\ie\ the QES is almost on the same null-time ($x^+$) slice as the
observer, but just a little \emph{towards} the null singularity\
(using absolute values gives\
$|x^+| - |x^+_0| \sim -{2Gc\over 3k\phi_r} |x^+_0|^{1-k}$).  The
location of the QES as being towards the singularity rather than away
as in the spacelike cases may look surprising at first sight.  However
from Figure~\ref{nullQES}, drawing constant $x^+$ and $x^-$ slices, it
is clear that the location of the QES with $\Delta x^+>0$ and $\Delta
x^-\ra-\infty$ is geometrically reasonable and expected if the QES and
the observer are to be spacelike separated\ ($\Delta x^+<0$ gives
timelike separation between the QES and the observer). In terms of
the $(t,r)$-coordinates (\ref{nullBgnd}), Figure~\ref{nullQES}
can be taken to depict the region with $x^+=t+r<0$ in the
$(t,r)$-plane, with the singularity locus being $t+r=0$ and the
timelike observer worldline having some fixed $r_0$ with $t_0<0$.
The description in Figure~\ref{nullQES} continues to hold as long 
as the observer remains timelike: it also holds if the observer is
moving along a null trajectory along $x^+$ with fixed $x^-$.
As a further check, we see that this extremization exhibits
time-maximization with null time ($x^+$): we have, using (\ref{nullQESext}),
\be
\del_+^2S_{gen} = -k(1-k){\phi_r\over 4G} (-x^+)^{k-2}
- {c\over 6} {1\over (x^+-x_0^+)^2}\quad\ra\quad \del_+^2S_{gen}|_*<0\ .
\ee
Note however that\ $\del_-^2S_{gen}=-{c\over 6}{1\over(\Delta x^-)^2}\ra 0^-$
from (\ref{SgenExtNull}). This should not be surprising: the 
2-dim space here is flat and the absence of the bulk gravitational field
makes it quite different from $AdS$-like spaces\ (\eg\ an expression
like $S\sim \log r$ gives $\del_r^2S\sim -{1\over r^2}\ra 0^-$).

As examples of (\ref{SgenNull}), we see that for a nearly smooth space
\eg\ with $k=\epsilon\ll 1$, (\ref{nullQESext}) gives\
$x_*^+ \sim (1-{2Gc\over 3\epsilon\phi_r}) x^+_0$\,.\
The case $k={2\over 3}$ gives the cubic
\be\label{nullQESk=2/3}
x^+_*=-t^3\,:\qquad t^3 + {Gc\over \phi_r}\,t - |x_0^+| = 0\ ,
\ee
which can be shown to have one real root which satisfies $\Delta x^+>0$
and agrees with (\ref{nullQESx+}) in perturbation theory in $Gc$.
For generic $k$ values, recasting using $x^+=-y^{{1\over 1-k}}$\,,
it can be seen numerically that there is one real root satisfying
$\Delta x^+>0$. Along these lines, for values such as $k={1\over 2}$
we choose the positive root of the resulting quadratic in continuity
with neighbouring $k$ values, which then again gives $\Delta x^+>0$.



Note that these null cosmological singularities are somewhat different
from the spacelike ones: for instance the extremization
(\ref{nullQESext}) shows that the singularity locus $x^+=0$ is in fact
an allowed QES solution when $x_0^+=0$. The behaviour near $x^+=0$ can
be seen explicitly in examples including (\ref{nullQESk=2/3}), \eg\
numerically. Thus these null singularities appear to not be excluded
from the entanglement wedge of the observer.\
However the on-shell generalized entropy (\ref{SgenNull}) continues
to be singular generically in the vicinity of the singularity:
(\ref{nullQESext}) gives
\be
S_{gen}^{o.s.} = {\phi_r\over 4G} (-x_*^+)^k + 
{c\over 6} \log \left({2Gc\over 3k\phi_r}\,
{(-x_*^+)^{1-k}\, |X_c^-|\over\epsilon_{UV}^2}\right)\ .
\ee
Thus although formal extrapolation to the singularity appears possible,
the above implies that the QES (\ref{nullQESext}) is only reliable in
the semiclassical regime with large $x^+_*$ and $Gc\ll 1$\ (where the
Bekenstein bound does not appear violated). Also since $S_{gen}^{o.s.}$
appears singular, further subleading contributions beyond the bulk
entropy term presumably also must be considered.
It was observed in \cite{Madhu:2009jh} that strings become highly
excited in the vicinity of a null Kasner Big-Crunch singularity\ (see
also \cite{Horowitz:1989bv,Craps:2008bv}). It is likely that this will
be true for (\ref{nullUpstairs2}) as well.  In this regard, note that
the backgrounds (\ref{nullBgnd}) necessarily require the extra scalar
$e^\Psi$ to be nontrivial: interpreting this as the string coupling
$g_s=e^\Psi$ and choosing the negative sign exponent for $e^\Psi$ in
(\ref{null2d}) suggests large string interactions in the vicinity of
the singularity $x^+=0$.  It is conceivable however that in some
appropriate double-scaling limit\ $x_*^+\ra 0,\ X_c^-\ra-\infty$,
with\ ${2Gc\over 3k\phi_r}\, {(-x_*^+)^{1-k}\, |X_c^-|\over\epsilon_{UV}}$\
held fixed, the generalized entropy can be rendered nonsingular. It
would be nice to explore this more carefully, perhaps dovetailing with
the positive sign exponent for $e^\Psi$ in (\ref{null2d}) and
suppressed string interactions. 

It is interesting to note that there is an entire function-worth of
nontrivial null backgrounds in (\ref{nullBgnd}), as (\ref{EOMnull})
shows. This is a special feature of 2-dim spacetimes that have a
``holomorphic'' structure, as is the case here with solely
$x^+$-dependence: for instance the backgrounds (\ref{nullUpstairs2})
can be recast by redefining the null-time variable to give
(\ref{nullBgnd}), so that the 2-dim metric is flat in these
$x^\pm$-coordinates\footnote{Instead of these ``flat'' variables, had
  we taken the background to be
\be
e^f=(X^+)^\al\,,\ \ \phi=(X^+)^K\,,\quad\ra\quad
(\del_+\Psi)^2 = {2K (\al-K+1)\over (X^+)^2}\ \ \ra\ \
0<K\leq \al+1\ .\nonumber
\ee
In other words, the exponent $k$ earlier is related as $k={K\over\al+1}$\,.
Now the generalized entropy contains the metric factor $e^{f/2}|_*$,
thus appearing singular.
}.
Spacelike cosmological singularities generically do not exhibit any
such ``holomorphy'' and cannot generically be recast in flat
coordinates and the metric factor $e^f$ lingers. This holomorphy
shows up in the extremization equations (\ref{SgenExtNull}),
(\ref{nullQESext}), where the $x^\pm$ sectors decouple (in contrast
with \eg\ (\ref{ExtSgen-r0t0}) in the AdS Kasner case, and more
generally (\ref{Sgen1})).  In fact, considering generic 2-dim
backgrounds (\ref{nullBgnd}), extremizing the generalized entropy
gives
\be
\del_+S_{gen} = 0 \quad \ra\quad
\del_+\phi + {2Gc\over 3\phi_r} {1\over x^+-x_0^+} = 0\quad\ra\quad
x^+ - x_0^+ = -{2Gc\over 3\phi_r} {1\over \del_+\phi}\ ,
\ee
again exhibiting this holomorphicity.
From the logic in Figure~\ref{nullQES} with $\Delta^2>0$ and\
$\Delta x^+>0$,\ $\Delta x^-\ra-\infty$, this implies that the
quantum extremal surface must lie in the direction of decreasing
dilaton, \ie\ $\del_+\phi<0$. This is consistent with our earlier
discussion since the dilaton Crunches towards decreasing $x^+$.

\section{Other cosmologies and QES}\label{sec:dSFRW}

In this section, we will study other cosmological backgrounds, in
particular de Sitter space in the Poincare slicing and FRW universes
under certain conditions.  One might take these to have natural
asymptotics at future or past timelike infinity, and are thus quite
different from the previous discussions on $AdS$-like or null
cosmologies where the asymptotics are at spatial or null infinity. As
we will see (and has been noted previously), the extremal surface
structure is rather different in these cases below: in some sense we
are simply extending our previous investigations in some formal way to
the cosmologies below, with the hope that better understanding will
emerge over time.

\subsection{de Sitter, Poincare}\label{sec:dSqes}

de Sitter space $dS_{d_i+1}$ in the Poincare slicing and its 2-dim
reduction are
\be
ds^2 = {R^2\over\tau^2}(-d\tau^2+dx^2+dy_i^2)\quad\ra\quad
\phi={R^{d_i}\over(-\tau)^{d_i}}\,,\ \ \
ds^2 = {R^{d_i+1}\over(-\tau)^{d_i+1}}(-d\tau^2+dx^2)\ .
\ee
We are parametrizing the upper Poincare patch with the future boundary
$I^+$ at $\tau=0$ and the past horizon at $\tau\ra-\infty$, and
$-\infty<\tau<0$ generically so the minus signs are explicitly retained.
As $\tau$ increases to the future, the dilaton grows. There is a
singularity at $\tau\ra-\infty$ in the effective 2-dim space: the
space is conformally $dS_2$\ (there are some parallels with
the discussions of $AdS_D$ reductions in \cite{Narayan:2020pyj}).

In this inflationary patch, we take the observer to be in the ground
state, so the bulk entropy is given by the ground state expression.
Then the generalized entropy for a bulk observer on a static worldline
at say $(x_0,\tau_0)$ is (see Appendix~\ref{App:EE2dCFT})
\be\label{SgendS}
S_{gen} = {\phi_r\over 4G}\,{R^{d_i}\over(-\tau)^{d_i}} +
{c\over 6}\log \left(\Delta^2\, {R^{(d_i+1)/2}\over(-\tau)^{(d_i+1)/2}}
\right)\ ,
\ee
retaining only terms relevant for extremization.
Then extremization gives
\be\label{SgendSext}
{c\over 3}\,{\Delta x\over\Delta^2} = 0\,,\qquad
- \left( -{d_i\phi_r\over 4G}\,{R^{d_i}\over(-\tau)^{d_i+1}}
- {c\over 12}\,{d_i+1\over(-\tau)}  \right)
- {c\over 3}\,{\tau-\tau_0\over \Delta^2} = 0\ .
\ee
One solution to this is
\be\label{dStimelikeQES}
\Delta x = 0\,,\ \ \Delta^2=-(\tau-\tau_0)^2\,;\ \qquad
{d_i\phi_r\over 4G}\,{R^{d_i}\over(-\tau)^{d_i+1}}
+ {c\over 12}\,{d_i+1\over(-\tau)} 
+ {c\over 3}\,{1\over\tau-\tau_0} = 0\ .
\ee
For a late time observer with $\tau_0\sim 0$, we have 
\be
\Delta x = 0\,,\quad
{d_i\phi_r\over 4G}\,{R^{d_i}\over(-\tau)^{d_i+1}}
+ {c\over 12}\,{3-d_i\over\tau} = 0
\ \ \ \ra\ \ \
\tau_*=-R\left({d_i\over 3-d_i}\,{3\phi_r\over Gc}\right)^{1/d_i}
\ee
Note that we are looking for a solution with $\tau<0$ as per our
parametrization: so for $d_i\geq 3$ (\ie\ $dS_5$ and higher) the
only real QES solution is $\tau\ra-\infty$.
For $d_i=1$ this matches the result in \cite{Chen:2020tes}\
((\ref{SgendS}) matches eq.6.7 there).

Most notably, the above QES solution is timelike-separated from the
observer: so $\Delta^2<0$\ (unlike \eg\ (\ref{t>t0}),
(\ref{nullQESext0})) and the generalized entropy (\ref{SgendS}) has
an imaginary part
from $\log(-1)$. In the spirit of our earlier discussions, it is
interesting to look for quantum extremal surfaces that are
spacelike-separated from the observer: towards this, note that there
is a distinct family of solutions which by construction are
spacelike-separated, along the lines of the discussions in the
cosmologies earlier with a regulator.  Then (\ref{SgendSext}) gives
\be\label{dSspacelikeQES}
\Delta^2\sim R_c^2\ ,\qquad 
{d_i\phi_r\over 4G}\,{R^{d_i}\over(-\tau)^{d_i+1}}
+ {c\over 12}\,{d_i+1\over(-\tau)} 
\sim {c\over 3}\,{\tau-\tau_0\over R_c^2}\ ,
\ee
regulating the QES as before with a spatial cutoff $R_c$.
First, note that if we remove the regulator so $R_c\ra\infty$, then
we obtain (with $t\equiv-\tau>0$)
\be
{d_i\phi_r\over 4G}\,{R^{d_i}\over t^{d_i+1}}
+ {c\over 12}\,{d_i+1\over t} = 0\ .
\ee
Both terms have the same sign so the only real QES is at $\tau\ra-\infty$.
This is spacelike separated only if the observer is also localized
at sufficiently early times.

With a finite spatial regulator $R_c$, we see that in general
$\tau>\tau_0$, \ie\ the QES lies on time slices later than the observer.
As a first approximation, note that in the classical limit $c\ra 0$,
the solution is $\tau\ra-\infty$\,: this is the location where the
dilaton is minimized. For early times also, the solution
is similar, \ie\
\be\label{dSrealQES}
|\tau_0|\gg R:\qquad \tau\ra-\infty\,,\ \ \tau\sim\tau_0\ ,
\ee
\ie\ the QES is in the far past when the observer is also in the
far past. This can be seen to exhibit time-maximization.
Let us analyze (\ref{dSspacelikeQES}) for 4-dim de Sitter upstairs
($d_i=2$): then
\be\label{dS4spqes}
\Delta^2\sim R_c^2\ ,\qquad
{\phi_r\over 2G}\,{R^{2}\over(-\tau)^{3}} + {c\over 12}\,{3\over(-\tau)}
\sim {c\over 6}\,{\tau-\tau_0\over R_c^2}\ .
\ee
With $t\equiv -\tau$, we can rewrite this as
\be
\Delta^2\sim R_c^2\ ,\qquad
{6\phi_r\over Gc}\,R^{2}R_c^2 + 3R_c^2\,t^2 \sim 2t^3 (t_0-t)\quad\ra\quad
t^4-t_0t^3+{3R_c^2\over 2}\,t^2+{3\phi_r\over Gc}\,R^2R_c^2 \sim 0\ .
\ee
Clearly as $t_0\ra 0$, there is no real QES solution since all terms are
positive (there appears to be a critical $t_0$ where the real QES
(\ref{dSrealQES}) stops existing). $dS_{d_i+1}$ can be seen to exhibit
similar behaviour.

Overall, in some essential sense, the physical interpretation of the
generalized entropy in these cases is not transparent, for instance as
holographic entanglement in the dual boundary theory, along the lines
of the $AdS$ cases\ (even from a bulk point of view alone, the
timelike separation is unconventional compared with the usual
formulations of entanglement on a spatial slice).  However our
discussion appears to corroborate previous work on classical de Sitter
extremal surfaces. Taking the future boundary as a natural anchor in
$dS$, there are either complex extremal surfaces
\cite{Narayan:2015vda,Sato:2015tta,Miyaji:2015yva} or future-past
(timelike) extremal surfaces
\cite{Narayan:2017xca,Narayan:2020nsc}. The latter future-past
surfaces perhaps suggest some new ``temporal entanglement'' between
$I^\pm$: taking the area of such surfaces to be real is effectively
removing an overall $i$-factor which would arise from rotating a
spatial extremal surface to a timelike one (this is also vindicated by
the complex generalized entropy (\ref{SgendS}), (\ref{dStimelikeQES}),
generalizing \cite{Chen:2020tes} for $dS_2$).  Overall perhaps this
suggests new interpretations towards entanglement in de Sitter space
based on the future boundary and $dS/CFT$
\cite{Strominger:2001pn,Witten:2001kn,Maldacena:2002vr}.  The $dS/CFT$
dictionary $\Psi_{dS}=Z_{CFT}$ suggests that boundary entanglement is
not bulk entanglement (quite unlike $Z_{bulk}=Z_{CFT}$ in $AdS$). Bulk
observables require $|\Psi_{dS}|^2$ suggesting two copies of the dual
CFT: this is reflected in the future-past extremal surfaces
\cite{Narayan:2017xca,Narayan:2020nsc} alluded to above. Other recent
perspectives on extremal surfaces anchored on the de Sitter horizon
include \eg\ \cite{Shaghoulian:2021cef}.

\subsection{FRW cosmologies, 2-dim gravity and QES}

Consider FRW cosmologies with flat spatial sections sourced by a
scalar field $\Psi$\ (general reviews include
\eg\ \cite{Trodden:2004st,Baumann:2009ds}): we choose one of the
spatial directions to be noncompact and perform dimensional reduction
on the others to obtain a 2-dim background 
\be
ds^2 = -dt^2 + a(t)^2 dx_i^2 \quad\ra\quad
\phi=a^{d_i}\,,\ \ \ ds^2 = a^{d_i+1} (-d\tau^2+dx^2)\ ,
\ee
as a solution to (\ref{actionXPsiU}), (\ref{2dimseom-EMD0-Psi}),
(\ref{2dimseom-EMD1-Psi}).
The energy-momentum conservation equation gives\
$dE+pdV=d(\rho\,a^{d_i+1})+pd(a^{d_i+1})=0$, \ie\
${\dot\rho}+(d_i+1)H(\rho+p)=0$.
This along with the Friedmann equation and the equation of state
$p=w\rho$ gives FRW cosmologies with
\be
p=w\rho\,,\quad a \sim t^k\,,\ \ k={2\over (1+d_i)(1+w)}
\qquad\
\Big[\rho={1\over 2}{\dot\Psi}^2 - V\,,\ \ p={1\over 2}{\dot\Psi}^2 + V\Big]
\ee
Now using conformal time $\tau$ gives
\be\label{ataunu}
d\tau = {dt\over a(t)}\ \ \ra\ \ \tau\sim t^{1-k}\quad\ra\quad
a(\tau) \sim \Big({\tau\over\tau_F}\Big)^{{k\over 1-k}}\equiv
\Big({\tau\over\tau_F}\Big)^\nu\ ,
\ee
introducing the FRW scale $\tau_F$ so the scale factor
becomes dimensionless: $\tau_F$ controls the strength of
time-dependence in these backgrounds, analogous to $t_K$ in
(\ref{AdSK-RtK}).
Note that the above FRW description is slightly different from
focussing on the vicinity of the singularity as in
\cite{Bhattacharya:2020qil}: taking dominant time derivatives implies
${\dot\Psi}^2\gg V$ so $p\sim \rho$, \ie\ $w\sim 1$, giving
$\nu={1\over d_i}$ so $\phi=a^{d_i}\sim \tau$ in agreement with the
universality (\ref{univSing}).
More generally the physical bounds on the equation of state parameter
$w$ translate to corresponding regimes for $\nu$:
\be\label{nu-w}
\nu = {2\over (1+d_i)(1+w) - 2}\ ;\qquad\ \  -1\leq w\leq 1\ \ \Rightarrow\ \
\nu>{1\over d_i}\ \ {\rm or}\ \ \nu\leq -1\ .
\ee

Now we analyse quantum extremal surfaces here. In general the bulk
matter entropy corresponds to some excited state, such as the thermal
state. A variety of such studies for FRW cosmologies including
entanglement with auxiliary universes appears in \cite{Hartman:2020khs},
revealing islands in various cases. Our discussion here will be
limited to simply extending the earlier de Sitter QES solutions to
certain FRW cases, which correspond to matter in the ground state
(as may arise for pressureless matter with $w=0$).
The generalized entropy for an observer in such a background is
(see Appendix~\ref{App:EE2dCFT})
\be\label{SgenFRW}
S_{gen} = {a^{d_i}\over 4G} + {c\over 6} \log
\left(\Delta^2\, a^{(d_i+1)/2}|_{(\tau,r)}\right)\ ,\qquad
\Delta^2 = (\Delta x)^2-(\tau-\tau_0)^2\ ,
\ee
where we are using conformal time $\tau$ in the 2-dim theory.
Now extremization gives
\bea\label{SgenFRWext}
\del_xS_{gen} = {c\over 6}{\del_x\Delta^2\over\Delta^2} = 0\ ,\ \ &&\ \
\del_\tau S_{gen} = {d_ia^{d_i-1}\,\del_\tau a\over 4G}
+ {c\over 12} {(d_i+1)\del_\tau a\over a}
+ {c\over 6} {\del_\tau\Delta^2\over\Delta^2} = 0\ ,\nonumber\\ [1mm]
\longrightarrow\qquad {c\over 3}\,{\Delta x\over\Delta^2} = 0\ ,\ \ &&\ \
{d_i\nu\over 4G} {\tau^{\nu d_i-1}\over\tau_F^{\nu d_i}}
       + {c\over 12} {(d_i+1)\nu\over\tau}
- {c\over 3} {\tau-\tau_0\over\Delta^2} = 0\ ,
\eea
using (\ref{ataunu}), (\ref{nu-w}). 
First considering timelike separated QES, we have
\be\label{FRWtimelikeQES}
   \Delta x = 0\ ,\ \ \Delta^2=-(\tau-\tau_0)^2\,;
   \qquad
       {d_i\nu\over 4G} {\tau^{\nu d_i-1}\over\tau_F^{\nu d_i}}
       + {c\over 12} {(d_i+1)\nu\over\tau}
+ {c\over 3} {1\over\tau-\tau_0} = 0\ ,
\ee
analogous to (\ref{dStimelikeQES}) in the de Sitter case $\nu=-1$:
for $\nu<-1$ the nature of these timelike QES is similar.
For $\nu d_i>1$, taking the first term to be dominant over the second
gives\
\be
\nu>{1\over d_i}\,:\qquad\quad
    {d_i\nu\over 4G} {\tau^{\nu d_i-1}\over\tau_F^{\nu d_i}}
+ {c\over 3} {1\over\tau-\tau_0} \sim 0\ \qquad\quad [\tau\gtrsim\tau_F]\ ,
\ee
which is structurally similar to (\ref{nullQESext0}), with
corresponding QES solutions (with $\tau_*-\tau_0<0$), valid for $\tau$
large compared to $\tau_F$. Since these are timelike-separated, the
on-shell generalized entropy acquires an imaginary part from $\log
(-1)$ in $\Delta^2<0$, similar to (\ref{dStimelikeQES}).

Alternatively, looking for spacelike separated QES along the lines of
(\ref{dSspacelikeQES}) gives
\be
\Delta^2\sim R_c^2\ , \qquad
      {d_i\nu\over 4G} {\tau^{\nu d_i-1}\over\tau_F^{\nu d_i}}
+ {c\over 12} {(d_i+1)\nu\over\tau} \sim {c\over 3} {\tau-\tau_0\over R_c^2}\ .
\ee
We are looking in the region of slow time evolution \ie\ large
$\tau\gtrsim\tau_F$\ (far from the singularity at $\tau=0$), towards
understanding the evolution of the QES with the observer time $\tau_0$.
Then for any $\nu>0$, we have $\tau^{\nu-1}>\tau^{-1}$ so we can
approximate the time extremization equation as
\be
   {d_i\nu\over 4G} {\tau^{\nu d_i-1}\over\tau_F^{\nu d_i}}
   \sim {c\over 3} {\tau-\tau_0\over R_c^2}
\quad\ra\quad
    \tau^{\nu d_i-1} \sim
    {4Gc\over 3d_i\nu}\,{\tau_F^{\nu d_i}\over R_c^2}\ (\tau-\tau_0).
\ee
This equation while tricky in general does have solutions at least
for specific families of $\nu$. For instance pressureless dust has
$w=0$ so using (\ref{nu-w}) we have
\bea
w=0\ \ \ie\ \ \nu={2\over d_i-1} &\xrightarrow{d_i=2} &
\tau^3\sim {Gc\,\tau_F^4\over 3 R_c^2}\ (\tau-\tau_0)\ ,\nonumber\\
&\xrightarrow{d_i=3} &
\tau^2\sim {4Gc\,\tau_F^3\over 9 R_c^2}\ (\tau-\tau_0)\ ,
\eea
both of which admit real solutions as long as
${Gc\tau_F^{\nu d_i}\over R_c^2}$ lies in appropriate regimes with
regard to $\tau_0$. For instance the $d_i=3$ case requires
${4Gc\,\tau_F^3\over 9 R_c^2}>4\tau_0$ for reality. Since the spatial
regulator $R_c\ra\infty$ strictly speaking, it is clear that these
solutions only make sense in a limit where we take $c$ small and
$R_c$ large holding the above condition fixed: so the existence of
these spacelike-separated QES solutions is not generic.

For generic scalar configurations, it is more appropriate
to consider bulk entropy contributions that are not those pertaining
to the ground state: then\ $S_{gen} = {a^{d_i}\over 4G} + S_b$\ gives\
${d_i\,\del_\tau a\over 4G} + \del_\tau S_b = 0$.
Discussions of this sort have previously appeared in \eg\ 
\cite{Hartman:2020khs}. When $S_b$ overpowers the classical area term,
the Bekenstein bound is violated and islands can arise if further
conditions hold. For $S_b$ representing bulk matter in some mixed
state, one might imagine some auxiliary purifying universe ``elsewhere''
which could then lead to islands. We will not discuss this further here.


\section{Discussion}\label{sec:Disc}

We have discussed quantum extremal surfaces in various cosmological
spacetimes with Big-Crunch singularities, developing further the
investigations in \cite{Manu:2020tty}. The generalized entropy here is
studied in 2-dimensional cosmologies obtainable in part from
dimensional reduction of higher dimensional cosmologies: the bulk
matter is taken to be in the ground state, which is reasonable in the
semiclassical region far from the singularity. First we focussed on
the isotropic $AdS$ Kasner spacetime and its reduction to
2-dimensions: the quantum extremal surfaces in \cite{Manu:2020tty}
were found to be driven to the semiclassical region infinitely far
from the Big-Crunch singularities present in these backgrounds (the
classical RT/HRT surfaces for finite subregion size bend in the
direction away from the singularity, Figure~\ref{cosRT1}). Analysing
further, the spatial extremization equation (\ref{rExt-reg}) shows
that in the semiclassical region, the QES location leads to the entire
Poincare wedge, with no island-like regions. Introducing a spatial
regulator in the time extremization equation (\ref{tExt-reg}) enables
understanding the dependence of the QES on the observer's location in
time.  This shows that the QES lags behind the observer location, in
the direction away from the singularity, as in
Figure~\ref{cosQES3}. The lag can be seen to increase slowly as the
observer evolves towards the singularity: extrapolating shows that the
singularity $t=0$ is not a solution to the extremization equations.
Thus the entanglement wedge appears to exclude the near singularity
region. Removing the regulator recovers the results in
\cite{Manu:2020tty}.  The spatial extremization equation
(\ref{rExt-reg}) shows an island-like region emerging for
(\ref{islandBndry}). However analysing carefully the extremization
equations recast as (\ref{Deltat/r-3}), (\ref{Dtr-tEx-quadr}), in the
vicinity of this island boundary reveals that the potential
island-like solution is in fact inconsistent.  Appending a
time-independent far region joined with the $AdS$ Kasner region at the
time slice $t=t_K$ as in (\ref{AdSK-RtK-reg}) gives further insight on
the QES behaviour. This QES analysis in the $AdS$ Kasner case extends
to more general singularities admitting a holographic interpretation,
with similar QES behaviour (\ref{rExt-reg-gen}), (\ref{tExt-reg-gen}),
in the semiclassical region, and inconsistencies near a potential
island boundary (\ref{islandBndry-gen}). These cosmologies include
nonrelativistic asymptotics: the assumptions on the exponents
(\ref{genCos}) are fairly general.

In sec.~\ref{sec:null}, we studied certain families of null Big-Crunch
singularities, which exhibit a certain ``holomorphy'' due to special
properties of null backgrounds. These are distinct in the behaviour of
the quantum extremal surface Figure~\ref{nullQES}, which can now reach
the singularity: however the on-shell generalized entropy continues to
be singular so the vicinity of singularity is not reliable.  In all
these cases, the QES is manifestly spacelike-separated from the
observer (\eg\ (\ref{t>t0}), (\ref{nullQESext0})), consistent with its
interpretation as holographic entanglement. We then discuss aspects of
2-dimensional effective theories involving dimensional reduction of
other cosmologies including de Sitter space (Poincare slicing) and FRW
cosmologies. In these cases, there are families of QES solutions which
are timelike-separated from the observer (\ref{dStimelikeQES}),
(\ref{FRWtimelikeQES}) \ (the $dS$ case here is in part a
generalization of some results in \cite{Chen:2020tes} for the $dS_2$
case): correspondingly the generalized entropy acquires an imaginary
part. We also find real spacelike-separated QES solutions in the
presence of finite spatial regulators (\ref{dSspacelikeQES}). In de
Sitter, these real solutions cease to exist for the late-time
observer.  Overall this perhaps corroborates earlier studies of
classical extremal surfaces anchored at the future boundary
\cite{Narayan:2015vda,Sato:2015tta,Miyaji:2015yva,
  Narayan:2017xca,Narayan:2020nsc}: see the discussion at the end
of sec.~\ref{sec:dSqes}.

Our investigations here have been on using quantum extremal surfaces
to gain some insights on cosmological spacetimes containing Big-Crunch
singularities: all these admit the form of a 2-dimensional cosmology
and thus exclude more general cosmologies that do not admit a
reduction to 2-dimensions. Most of our discussions pertain to bulk
matter in the ground state, which is reasonable far from the
singularities in the cosmologies we have discussed.  Overall the
cosmologies we have considered are closed universes with no horizons,
no appreciable entropy and no additional non-gravitating bath regions:
in such cases islands are not generic\ (there are parallels with some
discussions in \cite{Geng:2021hlu}). This is consistent with previous
studies of closed universes with no entanglement with ``elsewhere'',
\ie\ regions external to the universes in question which might act as
purifiers for mixed states. This is consistent with the Bekenstein
bound not being violated, \ie\ the bulk entropy does not overpower the
classical area in the generalized entropy.  Our discussion of de
Sitter space pertains only to the Poincare slicing: see
\eg\ \cite{Hartman:2020khs,Sybesma:2020fxg} for other discussions of
de Sitter. The FRW discussions also must be extended to cases with
bulk matter in excited states far from the ground state: in
this case islands will appear, corresponding to violations of the
Bekenstein bound.

Perhaps the most interesting question pertains to studying more
interesting models for bulk matter in the near-singularity spacetime
region where the matter might be expected to get highly excited.
Presumably incorporating analogs of more ``stringy'' or quantum
entanglement will give more insights into how the near singularity
region is accessible via entanglement (with the null singularities
perhaps more tractable).

At a more broad brush level, in some essential ways, cosmological
singularities in holography are perhaps qualitatively different from
black holes. They appear to require nontrivial non-generic initial
conditions: generic time-dependent deformations of the CFT vacuum are
expected to thermalize on long timescales, leading to black hole
formation in the bulk rather than a Big-Crunch. This appears
consistent with our finding that \eg\ the $AdS$ Kasner and other
holographic cosmological singularities are inaccessible via
entanglement with conventional ground state bulk matter: perhaps this
corroborates the expectation of non-generic holographic dual
states\ (see discussion after (\ref{AdSDK-2d}) and also other related
studies \eg\ \cite{Barbon:2015ria,Caputa:2021pad} of such
singularities and complexity).  It would be interesting to gain more
insights into the role of holographic entanglement, quantum extremal
surfaces and islands in cosmology more broadly.

\vspace{10mm}

{\footnotesize \noindent {\bf Acknowledgements:}\ \ It is a pleasure
  to thank Dileep Jatkar and A. Manu for comments on a draft. We also
  thank Debangshu Mukherjee for some discussions on the FRW cases.
  This work is partially supported by a grant to CMI from the Infosys
  Foundation.}

\vspace{5mm}

\appendix

\section{Some details: 2-dim gravity,\ extremal surfaces}\label{App:2dgES}

The equations of motion following from the 2-dim effective action
(\ref{actionXPsiU}) are
\bea\label{2dimseom-EMD0-Psi}
g_{\mu\nu}\nabla^2\phi-\nabla_{\mu}\nabla_{\nu}\phi
  +\frac{g_{\mu\nu}}{2}\Big(\frac{\phi}{2}(\partial\Psi)^2+U\Big)
  -\frac{\phi}{2}\partial_{\mu}\Psi\partial_{\nu}\Psi = 0\ ,\qquad\quad &&
  \nonumber\\ [1mm]
\mathcal{R}-\frac{\partial U}{\partial\phi}
  -\frac{1}{2}(\partial\Psi)^2 = 0\ ,\qquad\qquad
\frac{1}{\sqrt{-g}}\partial_{\mu}(\sqrt{-g}\,\phi \partial^{\mu}\Psi)
  -\frac{\partial U}{\partial\Psi} = 0\ .&&
\eea
In conformal gauge $g_{\mu\nu}=e^f\eta_{\mu\nu}$ these give
\bea\label{2dimseom-EMD1-Psi}
(tr)&& \qquad  \del_t\del_r\phi 
- {1\over 2} f'\del_t\phi - {1\over 2} {\dot f} \del_r\phi
+ {\phi\over 2}{\dot\Psi} \Psi' = 0\ , \nn\\
(rr+tt) && \ \ \
- \del_t^2\phi - \del_r^2\phi + {\dot f} \del_t\phi + f' \del_r\phi
- {\phi\over 2} ({\dot\Psi})^2 - {\phi\over 2} (\Psi')^2 = 0 ,\qquad \nn\\
[1mm]
(rr-tt) && \ \ \
- \del_t^2\phi + \del_r^2\phi + e^f U = 0\ ,\\ [1mm]
(\phi)&& \quad\  \big( {\ddot f} - f'' \big)
- {1\over 2} (-({\dot\Psi})^2+(\Psi')^2)
- e^f \frac{\partial U}{\partial\phi} = 0 ,\nn \\
(\Psi)&& \quad\   - \del_{t}(\phi\del_t\Psi) + \del_{r}(\phi\del_r\Psi)
- e^f {\del U\over\del\Psi} = 0\ . \nn 
\eea
The severe (singular) time-dependence in
the vicinity of the singularity implies that time-derivative terms are
dominant while other terms, in particular pertaining to the dilaton
potential, are irrelevant there: solving these leads to a
``universal'' subsector
\be\label{univSing}
\phi \sim t ,\quad\ e^f\sim t^a ,\quad\ e^\Psi\sim t^\al ;\qquad 
a={\al^2\over 2}\ ,
\ee
which governs the cosmological singularity. Analysing these equations
in more detail can be done using the ansatz (\ref{phie^fPsi-ansatz}),
giving \eg\ the $AdS$ Kasner cosmology (\ref{AdSDK-2d}) as well as
various others, some of which have nonrelativistic (hyperscaling
violating Lifshitz) asymptotics. For instance, flat space has $U=0$,
giving
\be\label{flat-2d}
\phi=t ,\quad ds^2=t^{\al^2/2} (-dt^2+dr^2) ,\quad e^\Psi=t^\al\ .
\ee
With $t=T^{1-p_1}$, these are the reduction of ``mostly
isotropic'' Kasner singularities\
$ds^2=-dt^2+t^{2p_1}dx_1^2+t^{2p_2}\sum_idx_i^2$.\ 
Hyperscaling violating cosmologies comprise backgrounds
(\ref{phie^fPsi-ansatz}) with exponents and parameters:
\bea\label{hvL-2d}
&& U(\phi,\Psi)=2\Lambda\phi^{{1\over d_i}} e^{\gamma\Psi}\,,\quad
\Lambda = -{1\over 2}(d_i+1-\theta)(d_i-\theta) ,\quad
\gamma = {-2\theta\over\sqrt{2d_i(d_i-\theta)(-\theta)}}\ , \nn\\
&& \qquad
m=-(d_i-\theta)\ ,\quad b={m(1+d_i)\over d_i}\ ,\quad \beta=-m\gamma\,,\nn\\
&& \qquad k=1 ,\quad a={\al^2\over 2}\,,\quad
\al=-\gamma \pm \sqrt{\gamma^2+{2(d_i-1)\over d_i}}\ .
\eea
Here $\theta<0$, $\gamma>0$. The higher dimensional backgrounds here
can be obtained as certain kinds of cosmological deformations of
reductions of nonconformal branes down to $D$ dimensions.\\
Still more complicated hyperscaling violating Lifshitz
cosmologies (with nontrivial Lifshitz exponents $z$ as well) and
their 2-dimensional avatars were also obtained in
\cite{Bhattacharya:2020qil}: these have a more complicated dilaton
potential. These are more constrained, requiring the conditions\
$m=-1,\ a=-b-2$, as well as further relations between other
exponents. A simple example has\ $\theta=0,\ z=2,\ d_i=2$,\ and\
$k=1,\ m=-1,\ a={1\over 2}\,,\ b=-{5\over 2}\,,\ \beta=-\al=1$,\
and the dilaton potential is\ $U=\phi^{1/2} (-3 + {1\over \phi^2} e^{-2\Psi})$.

\bigskip

\noindent \underline{\emph{Extremal (RT/HRT) surfaces:}}\ \
The area functional
\be
S = {V_{d_i-1}\over 4G_{d_i+2}} \int dr\, \phi\,
\sqrt{{e^f\over \phi^{(d_i+1)/d_i}}\big(1-(\partial_{r}t)^{2}\big)
  +(\partial_{r}x)^{2}}
\ee
upon extremizing $x(r)$ gives
\be\label{extSurf-Xf}
(\del_rx)^2 = A^2\,{{e^f\over \phi^{(d_i+1)/d_i}}\big(1-(\partial_{r}t)^{2}\big)
  \over \phi^2-A^2}\ ,\qquad
S = {V_{d_i-1}\over 4G_{d_i+2}} \int dr\; 
{e^{f/2}\,\phi^{(3-1/d_i)/2}\over \sqrt{\phi^2-A^2}}\, \sqrt{1-(\del_rt)^2}\ \ .
\ee
In the above expressions, $A$ is the turning point
$A=\phi_*={t_*\over r_*^{d_i}}$ for the $AdS$ Kasner case
(\ref{AdSDK-2d}). Analysing these extremal surfaces is reliable in the
semiclassical region far from the singularity at $t=0$. In this
region, a detailed analysis of the time extremization equation leads
to (\ref{t(r)Expn}): the surface lies almost on a constant time slice
($t''\ll 1$) and can be shown to bend in the direction away from the
singularity, as depicted in Figure~\ref{cosRT1}.

\section{Some details on 2d CFT and entanglement entropy}\label{App:EE2dCFT}

Any 2-dim metric is conformally flat so $ds^2=e^f\eta_{\mu\nu}dx^\mu dx^\nu$.
We can then modify the Calabrese-Cardy result
\cite{Calabrese:2004eu,Calabrese:2009qy}, in particular taking the
ground state entanglement in flat space and then incorporating the
effects of the conformal transformation $e^f$ as in \cite{Almheiri:2019psf}.
The twist operator 2-point function scales under a conformal
transformation as 
\be
\lan \sigma(x_1)\,\sigma(x_2)\ran_{e^fg} = e^{-\Delta_n\,f/2}\vert_{x_1}\,
e^{-\Delta_n\,f/2}\vert_{x_2}\, \lan \sigma(x_1)\,\sigma(x_2)\ran_{g}\ ,\qquad
\Delta_n={c\over 12} {n^2-1\over n}\ .
\ee
Since the partition function in the presence of twist operators scales
as the twist operator 2-point function, the entanglement entropy becomes
\be
S^{12}_{e^fg} = -\lim_{n\ra 1}\,\del_n \lan \sigma(x_1)\,\sigma(x_2)\ran_{e^fg}
= S^{12}_g + {c\over 6}\sum_{endpoints} \log e^{f/2}\ .
\ee
For a bulk interval, this gives
\be
S^{12}_g = {c\over 6}\log \left({\Delta^2\over\epsilon_{UV}^2}\right)
\quad\ra\quad
S^{12}_{e^f g} = {c\over 6}\log \left({\Delta^2\over\epsilon_{UV}^2}\,
e^{f/2}|_1\, e^{f/2}|_2\right)
\ee
while for a CFT with boundary, we have essentially half the flat
space answer (with one end of the interval at the boundary), thus obtaining
\be
S^{10}_g = {c\over 12} \log\left({\Delta^2\over\epsilon_{UV}^2}\right)
\quad\ra\quad
S^{10}_{e^f g} = {c\over 12}\log \left({\Delta^2\over\epsilon_{UV}^2}\,
e^f|_1 \right)
\ee
We have used the latter in the $AdS$ cases which include the presence
of the $AdS$ boundary, while for the bulk cases we use the former
expression.


\vspace{7mm}

\end{document}